\RequirePackage[hyphens]{url}
\documentclass[reqno]{article}

\usepackage[utf8]{inputenc}  
\usepackage[T1]{fontenc}    
\usepackage[english]{babel}
\usepackage{amsmath,amsthm} 
\usepackage{amssymb,mathrsfs} 
\usepackage{makeidx}
\usepackage{breakcites}
\usepackage{amsfonts}
\usepackage{nicefrac}
\usepackage{geometry}
\usepackage{authblk}
\usepackage{graphicx}
\usepackage{color}
\usepackage{oldgerm}
\usepackage{relsize}
\usepackage{mathrsfs}
\usepackage{stmaryrd}
\usepackage{enumitem}
\usepackage[active]{srcltx}
\usepackage{verbatim}
\usepackage{lmodern} 
\usepackage{xcolor}
\usepackage[toc,page]{appendix}
\usepackage{aliascnt,bbm}
\usepackage{array}
\usepackage[colorlinks = true, citecolor = blue, breaklinks=true]{hyperref}
\usepackage{breakurl}
\usepackage[textwidth=3cm,textsize=footnotesize]{todonotes}
\usepackage{bbm}
\usepackage{ifthen}
\usepackage{xargs}
\usepackage{cellspace}
\usepackage[boxed,noline]{algorithm2e}
\makeatletter
\renewcommand{\@algocf@capt@boxed}{above}
\makeatother
\SetAlgoNoLine
\usepackage{array,multirow,makecell}
\usepackage[Symbolsmallscale]{upgreek}
\usepackage{natbib}
\usepackage{accents}
\usepackage{dsfont}
\usepackage{aliascnt}
\usepackage{xr}
\usepackage{cleveref}

\makeatletter
\newtheorem{theorem}{Theorem}
\crefname{theorem}{theorem}{Theorems}
\Crefname{Theorem}{Theorem}{Theorems}
\newtheorem*{lemma_nonumber*}{Lemma}

\newaliascnt{lemma}{theorem}

\aliascntresetthe{lemma}
\crefname{lemma}{lemma}{lemmas}
\Crefname{Lemma}{Lemma}{Lemmas}

\newaliascnt{corollary}{theorem}

\aliascntresetthe{corollary}
\crefname{corollary}{corollary}{corollaries}
\Crefname{Corollary}{Corollary}{Corollaries}

\newaliascnt{proposition}{theorem}

\aliascntresetthe{proposition}
\crefname{proposition}{proposition}{propositions}
\Crefname{Proposition}{Proposition}{Propositions}

\newaliascnt{definition}{theorem}

\aliascntresetthe{definition}
\crefname{definition}{definition}{definitions}
\Crefname{Definition}{Definition}{Definitions}

\newaliascnt{remark}{theorem}

\aliascntresetthe{remark}
\crefname{remark}{remark}{remarks}
\Crefname{Remark}{Remark}{Remarks}

\newtheorem{example}[theorem]{Example}
\crefname{example}{example}{examples}
\Crefname{Example}{Example}{Examples}

\crefname{figure}{figure}{figures}
\Crefname{Figure}{Figure}{Figures}

\crefformat{assumption}{{\textbf{A}}#2#1#3}

\Crefname{assumptionG}{\textbf{G}\hspace{-3pt}}{\textbf{G}\hspace{-3pt}}
\crefname{assumptionG}{\textbf{G}}{\textbf{G}}

\Crefname{assumptionQ}{\textbf{Q}\hspace{-3pt}}{\textbf{Q}\hspace{-3pt}}
\crefname{assumptionQ}{\textbf{Q}}{\textbf{Q}}

\def\bfe{\mathbf{e}}

\def\msa{\mathsf{A}}

\def\mse{\mathsf{E}}

\def\msv{\mathsf{V}}

\def\msy{\mathsf{Y}}

\def\mcbb{\mathcal{B}}  
\newcommand{\mcb}[1]{\mathcal{B}(#1)}

\def\rset{\mathbb{R}}

\def\nset{\mathbb{N}}

\def\mrc{\mathrm{C}}

\def\rmd{\mathrm{d}}
\def\rmT{\mathrm{T}}
\def\rmC{\mathrm{C}}
\def\rmO{\mathrm{O}}
\def\mrd{\mathrm{d}}

\def\rme{\mathrm{e}}
\def\rmn{\mathrm{n}}
\newcommand{\rmna}[1]{\mathrm{n}\parenthese{#1}}
\def\mrn{\mathrm{n}}

\newcommandx{\functionspace}[2][1=+]{\mathbb{F}_{#1}(#2)}

\newcommand{\argmin}{\operatorname*{arg\,min}}

\newcommandx{\VarDeux}[3][3=]{\operatorname{Var}^{#3}_{#1}\left\{#2 \right\}}

\newcommand{\1}{\mathbbm{1}}

\newcommand{\LeftEqNo}{\let\veqno\@@leqno}

\newcommand{\floor}[1]{\left\lfloor #1 \right\rfloor}

\newcommand{\N}{\ensuremath{\mathbb{N}}}

\newcommand{\abs}[1]{\left\vert #1 \right\vert}

\newcommandx{\Vnorm}[2][1=V]{\| #2 \|_{#1}}
\newcommandx{\VnormEq}[2][1=V]{\left\| #2 \right\|_{#1}}
\newcommandx{\norm}[2][1=]{\ifthenelse{\equal{#1}{}}{\left\Vert #2 \right\Vert}{\left\Vert #2 \right\Vert^{#1}}}
\newcommandx{\normLigne}[2][1=]{\ifthenelse{\equal{#1}{}}{\Vert #2 \Vert}{\Vert #2\Vert^{#1}}}

\newcommand{\parenthese}[1]{\left(#1 \right)}
\newcommand{\parentheseLigne}[1]{(#1 )}
\newcommand{\parentheseDeux}[1]{\left[ #1 \right]}
\newcommand{\defEns}[1]{\left\lbrace #1 \right\rbrace }

\newcommand{\ps}[2]{\left\langle#1,#2 \right\rangle}
\newcommand{\psLigne}[2]{\langle#1,#2 \rangle}

\newcommand{\proba}[1]{\mathbb{P}\left( #1 \right)}

\newcommandx\probaMarkovTilde[2][2=]
{\ifthenelse{\equal{#2}{}}{{\widetilde{\mathbb{P}}_{#1}}}{\widetilde{\mathbb{P}}_{#1}\left[ #2\right]}}

\newcommand{\plusinfty}{+\infty}

\def\ie{\textit{i.e.}}

\def\eqsp{\;}
\newcommand{\coint}[1]{\left[#1\right)}

\newcommand{\ooint}[1]{\left(#1\right)}
\newcommand{\ccint}[1]{\left[#1\right]}

\newcommandx{\weight}[2][2=n]{\omega_{#1,#2}^N}

\newcommandx\sequence[3][2=,3=]
{\ifthenelse{\equal{#3}{}}{\ensuremath{\{ #1_{#2}\}}}{\ensuremath{\{ #1_{#2}, \eqsp #2 \in #3 \}}}}
\newcommandx\sequenceD[3][2=,3=]
{\ifthenelse{\equal{#3}{}}{\ensuremath{\{ #1_{#2}\}}}{\ensuremath{( #1)_{ #2 \in #3} }}}

\newcommandx{\sequencen}[2][2=n\in\N]{\ensuremath{\{ #1_n, \eqsp #2 \}}}
\newcommandx\sequenceDouble[4][3=,4=]
{\ifthenelse{\equal{#3}{}}{\ensuremath{\{ (#1_{#3},#2_{#3}) \}}}{\ensuremath{\{  (#1_{#3},#2_{#3}), \eqsp #3 \in #4 \}}}}
\newcommandx{\sequencenDouble}[3][3=n\in\N]{\ensuremath{\{ (#1_{n},#2_{n}), \eqsp #3 \}}}

\def\iid{i.i.d.}

\def\eg{e.g.}

\newcommand{\opnorm}[1]{{\left\vert\kern-0.25ex\left\vert\kern-0.25ex\left\vert #1 
    \right\vert\kern-0.25ex\right\vert\kern-0.25ex\right\vert}}

\def\generator{\mathcal{A}}

\def\Id{\operatorname{Id}}

\newcommandx{\CPE}[3][1=]{{\mathbb E}_{#1}\left[#2 \left \vert #3 \right. \right]} 
\newcommandx{\CPVar}[3][1=]{\mathrm{Var}^{#3}_{#1}\left\{ #2 \right\}}
\newcommand{\CPP}[3][]
{\ifthenelse{\equal{#1}{}}{{\mathbb P}\left(\left. #2 \, \right| #3 \right)}{{\mathbb P}_{#1}\left(\left. #2 \, \right | #3 \right)}}

\newcommandx{\osc}[2][1=]{\mathrm{osc}_{#1}(#2)}

\def\Id{\operatorname{Id}}
\def\IdM{\operatorname{I}_d}

\def\lambdab{\bar{\lambda}}
\def\blambda{\bar{\lambda}}

\def\transpose{\operatorname{T}}

\def\yprime{y'}

\def\bX{\bar{X}}
\def\bY{\bar{Y}}

\def\yt{\tilde{y}}
\def\ty{\yt}
\def\xt{\tilde{x}}
\def\tx{\xt}

\def\bB{\bar{B}}
\def\bQ{\bar{Q}}
\def\bP{\bar{P}}
\def\bvarphi{\bar{\varphi}}
\def\bgenerator{\bar{\mathcal{A}}}

\def\bX{\bar{X}}
\def\bY{\bar{Y}}

\def\sphere{\mathbb{S}}

\def\sign{\operatorname{sign}}
\def\rate{\bar{\lambda}}

\newcommand{\ensemble}[2]{\left\{#1\,:\eqsp #2\right\}}

\def\kernel{M}
\def\kernelQ{Q}
\def\Qkernel{Q}
\def\tpi{\tilde{\pi}}
\def\rate{\lambda}

\def\loiy{\mu_{\msy}}
\def\proj{\mathrm{\proj}}
\def\projp{\mathrm{P}^{x,\sslash}}
 
 \def\projn{\mathrm{P}^{x,\perp}}
  \newcommand\projnn[1]{\mathrm{P}^{#1,\perp}}
 \def\Kp{K^{x,\sslash}}
 \def\Kn{K^{x,\yp,\perp}}
 
 \def\tKn{\tilde{K}^{x,\perp}}
 \newcommand\Knn[2]{K^{#1,#2,\perp}}
 \newcommand\Kpp[1]{K^{#1,\sslash}}
  \newcommand\tKnn[1]{\tilde{K}^{#1,\perp}}

 \def\tKn{\tilde{K}^{x,\perp}}
  \def\Qp{Q^{x,\sslash}}
\def\Qn{Q^{x,\perp}}

\def\yp{y_{\sslash}}
\def\yn{y_{\perp}}
\def\Yn{Y_{\perp}}
\def\spana{\mathrm{span}}

\def\mrl{\mathrm{L}}
\def\div{\operatorname{div}}

\def\mupp{\mu_{\mathrm{p}}}
\def\loiyp{\loiy^{\mbox{\smaller[5]{$\sslash$}}}}
\def\loiym{\rho}
\newcommand{\fracp}[1]{\mathrm{f}\parenthese{#1}}
\def\matOrt{\mathbf{O}}

\def\Idd{\mathrm{I}_d}
\def\lambdatot{\lambda_{\mathrm{t}}}
\def\Qtot{Q_{\mathrm{t}}}
\def\generatortot{\generator_{\mathrm{t}}}
\def\diag{\mathrm{diag}}
\def\hh{\hat{h}}
\def\rmint{\mathrm{int}}
\def\ESS{\mathrm{ESS}}

\def\bfx{\mathbf{x}}
\def\xprime{x'}

\title{Forward Event-Chain Monte Carlo: Fast sampling by randomness control in irreversible Markov chains}

\author[1]{Manon Michel}
\author[2]{Alain Durmus}
\author[3]{St\'ephane S\'en\'ecal}

\affil[1]{Laboratoire de mathématiques Blaise Pascal, UMR 6620, CNRS, Université Clermont-Auvergne, 63718 Aubière Cedex, France.}
\affil[2]{CMLA, UMR 8536, \'Ecole normale supérieure Paris-Saclay, CNRS, Université Paris-Saclay, 94235 Cachan, France.}
\affil[3]{Orange Labs, 44 avenue de la R\'epublique, CS 50010, 92326 Chatillon Cedex, France}

\begin{document}
\maketitle
\footnotetext[1]{Email: manon.michel@uca.fr. This work was achieved while M. M. worked at Orange Labs and Centre de Mathématiques Appliquées, École Polytechnique.}
\footnotetext[2]{Email: alain.durmus@cmla.ens-cachan.fr}

\begin{abstract}
  Irreversible and rejection-free Monte Carlo methods, recently
  developed in physics under the name Event-Chain and known in
  Statistics as Piecewise Deterministic Monte Carlo (PDMC), have
  proven to produce clear acceleration over standard Monte Carlo
  methods, thanks to the reduction of their random-walk
  behavior. However, while applying such schemes to standard
  statistical models, one generally needs to introduce an additional
  randomization for sake of correctness. We propose here a new class
  of Event-Chain Monte Carlo methods that reduces this
  extra-randomization to a bare minimum. We compare the efficiency of
  this new methodology to standard PDMC and Monte Carlo
  methods. Accelerations up to several magnitudes and reduced
  dimensional scalings are exhibited.
\end{abstract}

\section{Introduction}
\label{sec:intro}

Markov Chain Monte Carlo (MCMC) algorithms are commonly used for the
estimation of complex statistical distributions
\citep{mcstat}. The core idea of these
methods is to design a Markov chain, whose invariant distribution is
the a posteriori distribution associated with a statistical model of
interest. However, naive MCMC methods, based on reversible
Markov chains, are often challenged by multimodal and
high-dimensional target distributions since they often display a
diffusive behavior and can be impeded by high rejection
rate. Important efforts have been devoted to the design of
non-reversible and rejection-free schemes, seeking the reduction of
the random-walk behavior.

The Hamiltonian dynamics used in Hybrid/Hamiltonian Monte Carlo
algorithms \citep{hymc,bayeslnn} provides an example of such
alternative frameworks \citep{hmc,adaphmc,nuts}. These methods require
however a fine tuning of several parameters, alleviated recently by
the development of the statistical software Stan
\citep{stan:2017}. Also, while aiming at introducing persistency in
the successive steps of the Markov chain, these methods still rely on
reversible chains with an acceptance-reject scheme.

In physics, recent advances were made in the field of irreversible and
rejection-free MCMC simulation methods. These new schemes, referred to
as Event-Chain Monte Carlo
\citep{bernard:2009,michel:kapfer:krauth:2014}, generalize the concept
of lifting developed by \cite{Diaconis2000}, while drawing on the
lines of the recent rejection-free Monte Carlo algorithm introduced in
\cite{peters:dewith:2012}. Their successes in different applications
\citep{Bernard_2011, Kapfer_2015} have motivated the development of a
general framework based on Piecewise Deterministic Markov Processes
(PDMP) and some numerical experiments show an acceleration in
comparison to the Hamiltonian MC, see
\cite{bouchard:vollmer:doucet:2016,bierkens:roberts:2016,bierkens:fearnhead:roberts:2016}. Nevertheless,
PDMC methods can still suffer from some random-walk behavior, partly
because they still rely on an additional randomization step to ensure
ergodicity.

In this paper, we introduce a generalized PDMC framework, the Forward
Event-Chain Monte Carlo. This method allows for a fast and global
exploration of the sampling space, thanks to a new lifting
implementation which leads to a minimal randomization and an
alleviation of critical parameters tuning. In this framework, the
successive directions are picked according to a full probability
distribution conditional on the local potential gradient, contrary to
previous PDMC.  This paper is organized as follows. \Cref{sec:mcmc}
first recalls and describes the standard MCMC sampling methodologies,
as well as classical PDMC sampling schemes. Then,
\Cref{sec:derivation-new-pdmps} introduces the original Forward
Event-Chain Monte Carlo framework method proposed in the
paper. \Cref{sec:numexp} illustrates the performances of the proposed
framework for high-dimensional ill-conditioned Gaussian distributions,
a Poisson-Gaussian Markov random field model, mixtures of Gaussian
distributions and logistic regression problems. Speedups of several
magnitudes in comparison to standard PDMC implementations are shown.

\section{Piecewise deterministic Markov processes for Monte Carlo methods}
\label{sec:mcmc}
\subsection{Towards irreversible MCMC Sampling}

We consider in this paper a target probability measure $\pi$ which
admits a positive density with respect to the Lebesgue measure $\pi :
\rset^d \to \rset_+^*$ of the form
$  \pi(x) =  \rme^{-U(x)} /\int_{\rset^d} \rme^{-U(\xprime)}
  \rmd \xprime $ for all $x \in \rset^d$,
where $U : \rset^d \to \rset$ is a continuously differentiable
function, referred to as the potential associated with $\pi$.  MCMC
sampling techniques are implemented through the recursive application
of a Markov kernel, denoted as $K$, such that $\pi$ is an invariant
distribution, \ie~$\pi K = \pi$, which is equivalent to 
\begin{equation}
   \int_{\xprime\in \rset^d} \pi(\rmd \xprime)K(  \xprime,  \rmd x) = \int_{\xprime \in \rset^d} \pi(\rmd x) K(x, \rmd \xprime) \eqsp,
\label{eq:GB}
\end{equation}
also known as the global-balance condition.

The most common approach to satisfy the relation \eqref{eq:GB} is to
consider the following sufficient stronger condition on $K$:
$ \pi(\rmd x')K( x', \rmd x) = \pi(\rmd x) K(x, \rmd x')$, referred
to as the detailed-balance (or reversibility) condition.
This condition enforces the artificial constraint of a local symmetry
between any two pairs of states $x,x'\in \rset^d$, which yields the
self-adjoint property of $K$ in $\mrl^2(\pi)$, the set of measurable functions $f : \rset^d\to \rset^d$ satisfying $\int_{\rset^d} f^2(\tilde{x}) \rmd \pi(\tilde{x}) < \plusinfty$. In most cases, it leads
to rejections and a random-walk behavior, which impede the sampling
efficiency. However this local symmetry allows for an easy construction of
general Markov kernels $K$ and thus played a large part in the
popularity of detailed-balance methods. Most prominent MCMC schemes
like the Hastings-Metropolis \citep{metro,hastings} and the Gibbs
sampling \citep{gibbs1,gibbs2} algorithms belongs to this class.

Irreversible or non-reversible MCMC samplers have attracted a lot of
attention for the last two decades. They break the detailed-balance
condition while still obeying the global-balance one and leaving $\pi$
invariant and, by doing so, have often been shown to have better
  convergence compared to their reversible counterpart. It is however
  still challenging to develop a construction methodology for
  irreversible kernels, which displays the generality of reversible
  schemes as the Metropolis-Hastings algorithm, while improving the
  convergence. Indeed, non-reversible MCMC algorithms can be directly
  built from the composition of reversible MCMC kernels
  (e.g. Deterministic Scan Gibbs samplers \cite{gelfand1990sampling}), but it is well-known that
  such a strategy can be relatively inefficient, in particular since
  it does not prevent diffusive behavior and backtracking in the
  resulting process. To circumvent this issue, popular solutions
  consist in extending the state space by introducing an additional
  variable and targeting the extended probability distribution
\begin{equation}
  \label{eq:extended_target}
  \tpi = \pi \otimes \loiy  \eqsp,
\end{equation}
where $\loiy$ is a probability distribution on $\rset^d$, endowed with
the Borel $\sigma$-field $\mcbb(\rset^d)$. First sampling from $\loiy$
and then fixing the proposal distribution accordingly allow for the
implementation of persistent moves. It has been shown that
non-reversible MCMC relying on such approaches can improve the
$\mathrm{L}^2$ spectral gap and the asymptotic variance of MCMC
estimators based on these methods, see
\eg~\cite{lifting,Diaconis2000,Neal2004}. Henceforth we refer to the
additional variable $y$ as the direction and $x$ as the position.

If this approach allows for general implementations, the
  challenge lies in finding a good direction update strategy, which
  preserves ergodicity and performs efficiently. Historically,
  such irreversible Markov chain samplers have been introduced under
  the name of \emph{Hybrid} or \emph{Hamiltonian} Monte Carlo in
  \cite{hymc} and the name of \emph{lifted Markov chains} in
  \cite{lifting} and \cite{Diaconis2000}. In the former, the proposal
  distribution follows a Newtonian dynamics, ergodicity is ensured
  through direction refreshment and correctness through rejections. In
  the latter, a direction is fixed over the state space, rejections
  are transformed into direction changes and ergodicity is ensured by
  a partition of the state space by direction lines.
Nevertheless, they all have in common to rely on a \emph{skew}
detailed-balance condition \citep{sakai:hukushima:2013}: while
formally breaking the detailed-balance condition, correctness is still
ensured by the following local symmetry condition,
$ \tpi(\rmd x',\rmd y')K( (x',y'), \rmd x \rmd y) = \tpi(\rmd x, \rmd
y) K((x,-y), \rmd x' \rmd y')$.
Lately, irreversible schemes violating also the skew detailed-balance
conditions have been developed in physics
\citep{peters:dewith:2012,michel:kapfer:krauth:2014}. They fix a
  direction and are rejection-free, as done in
  lifting schemes, but relies on a direction
  shuffling for ergodicity, as done in HMC. These methods are not
based on an artificial skew symmetry but on intrinsic ones of the
extended target distribution $\tpi$ itself. This idea was recently
developed in \cite{polymer} to sample from target distributions which
are assumed to be divergence-free,
\ie~$\div(U)= \sum_{i=1}^d \partial U / \partial x_i =0$, extending
the first methods of \cite{peters:dewith:2012} and
\cite{michel:kapfer:krauth:2014} which require factorizable
distributions of the form
$U(x) = \sum_{1 \leq i < j \leq d} U_{i,j}(x_i,x_j)$, where
$U_{i,j} : \rset^n \to \rset$ for all $i,j \in \{1,\ldots,d\}$,
satisfying the local divergence-free condition $\div(U_{i,j}) = 0$.
These schemes simulate ballistic trajectories over the state space,
whose direction changes at random times called events, forming up an
\emph{event chain}. They have been described as a piecewise
deterministic Markov process (PDMP) \citep{davis:1993}, as explained
in the next section, and adapted to a Bayesian setting in
\cite{bierkens:fearnhead:roberts:2016,bouchard:vollmer:doucet:2016}.

Ideally, one would like to find the optimal set of directions
  necessary for ergodicity and allowing for an efficient exploration
  and update the direction among this set at events. But, contrary to
  physics, many statistical models are not chosen based on some a
  priori knowledge or a basis allowing for an efficient factorization
  and it is not possible to rely on a sparse direction set. Also, in
  the absence of natural symmetry similar to the divergence-free
  condition, a local symmetry is again imposed by a deterministic
  change of directions, which comes down to a skew-detailed
  balance. Finally, to ensure ergodicity of some of these methods,
  e.g. BPS, the direction has to be resampled, partially or totally,
  according to $\loiy$ during the simulation and this refreshment has
  a direct impact on the asymptotic variance of the Monte Carlo
  estimators, see \eg~\cite{andrieu2019peskun}. Ergodicity can
  also be ensured by a line partition of the state space, as done
  originally in the lifting framework and is the case in the Zig-Zag
  (ZZ) scheme \citep{bierkens:roberts:zitt;2017}, which relies on a
  reduction of the multidimensional problem into a collection of
  unidimensional ones through factorization. This approach can lead to
  a slow exploration if the direction lines are not aligned on the
  target distribution and is relying on a collection of
  unidimensional skew-detailed balances.

The object of this paper is to show that such additional symmetry is
not needed to design general irreversible schemes and how to do
so. One of the key ideas is to rely on a stochastic picture by
considering the full probability distribution of the direction at the
events. Such randomized change of directions were first considered by
\cite{michel:2016} and \cite{bierkens:fearnhead:roberts:2016}, but
without specifying this general distribution and highlighting the role played
by the decomposition along the potential gradient. In addition, we
propose new refreshment strategies which, by being coupled to the
stochastic direction changes, reduce the amount of noise needed for
ergodicity and therefore limit the diffusive behaviour. We name this
generalized class of PDMC algorithms \emph{Forward} Event-Chain Monte
Carlo as the underlying process keeps on going forward, while breaking
free from local symmetry. We finally exhibit how the new degrees of
freedom of refreshment and direction changes of the Forward EC methods
can improve on existing PDMC methods and do not require any
fine tuning to be efficient.

In this paper, for the sake of clarity, we consider $\loiy$ to be
either the uniform distribution on
$\sphere^{d-1} = \{ y \in \rset^d \, : \, \norm{y} = 1\}$ or the
$d$-dimensional standard Gaussian distribution. However, the presented
methodology can be adapted to more general auxiliary distribution
$\loiy$.  In the sequel, we denote by $\msy$ the support of $\loiy$,
therefore $\msy$ is either $\sphere^{d-1}$ or $\rset^d$.

\subsection{Piecewise Deterministic Monte Carlo}
\label{sec:forward-p-orthogonal}
A PDMP $(X_t,Y_t)_{t\geq 0}$ is completely defined on $\rset^{2d}$ by
giving an initial state $(X_0, Y_0)$, a smooth deterministic
differential flow $(\varphi_t)_{t\geq 0}$, a Markov kernel on
$(\rset^{2d},\mathcal{B}(\rset^{2d}))$, denoted by $\kernel$ and a
function $\rate : \rset^{2d} \to \rset_+$, referred to hereinafter as
the event rate. The data $(\varphi,\kernel,\lambda)$ is called the
characteristics of the PDMP $(X_t,Y_t)_{t \geq 0}$.

The differential flow $(\varphi_t)_{t\geq 0}$ sets the evolution of
the process $(X_t,Y_t)_{t\geq 0}$ for $t \in \coint{S_n, S_{n+1}}$, as
$(X_t,Y_t) = \varphi_{t-S_n}(X_{S_n},Y_{S_n})$. The event times
$(S_k)_{k\in \nset}$ are defined recursively by $S_0=0$ and for
$n \geq 0$, $S_{n+1} = S_n + T_{n+1}$ where $T_{n+1}$ is a
$\bar{\rset}$-random variable independent of the past with survival
function $  \mathbb{P}(T_{n+1} \geq t) = \exp\defEns{- \int_{0}^{t}
    \rate(\varphi_s(X_{S_n}, Y_{S_n})) \rmd s}$, for all $t
  \geq 0$.
At an event time $S_{n+1}$, $(X_{S_{n+1}},Y_{S_{n+1}})$ is drawn from
the distribution $\kernel((X_{S_{n+1}-},Y_{S_{n+1}-}), \cdot)$ where
we set
$(X_{S_{n+1}-},Y_{S_{n+1}-}) = \varphi_{T_{n+1}}(X_{S_n},Y_{S_n})$.  We
assume the usual condition $\sup_{n\in\nset}S_n=\plusinfty$, that will
be satisfied in our application. Under appropriate conditions
\cite[Theorem 25.5]{davis:1993}, the process $(X_t,Y_t)_{t \geq 0}$ is
strongly Markovian.
 In addition, its probability distribution defines a Markov
semi-group $(P_t)_{t \geq 0}$ for all $(x,y) \in \rset^{2d}$ and
$\msa \in \mcb{\rset^{2d}}$, by $  P_t((x,y), \msa) = \proba{(X_t,Y_t) \in \msa}$,
where $(X_0, Y_0)=(x,y)$.

In the following, we consider the differential flow $(\varphi_t)_{t\geq 0}$ on
$\rset^d$ associated with the Ordinary Differential Equation (ODE),
$(\dot{x}_t, \dot{y}_t)= (y_t,0)$ and given for all $(x,y) \in
\rset^d$ and $t \geq 0$ by
\begin{equation}
\label{eq:differential_flow}
  \varphi_t(x,y) = (x+ty , y) \eqsp.
\end{equation}
This flow is the one used in most lifted MCMC schemes.
Regarding the event rate $\lambda$, we set
\begin{equation}
\label{eq:definition_rate}
\rate(x,y) = \ps{y}{\nabla U(x)}_+ \eqsp +\lambdab \eqsp, \text{ \qquad with $\lambdab\in\rset_+$} \eqsp,
\end{equation}
and a Markov kernel $\kernel$ of the
following form, for all $\mathsf{A} \in \mcbb(\rset^{2d})$ and $(x,y)
\in \rset^{2d}$,
\begin{equation}
\label{eq:definition_kernelKQ}
\kernel((x,y),\mathsf{A}) = \frac{\ps{y}{\nabla U(x)}_+}{\rate(x,y)} \, \int_{\msy }
  \1_{\mathsf{A}}(x,\ty) \kernelQ((x,y),\rmd \ty) + \frac{\lambdab}{\rate(x,y)}\loiy(\msa)  \eqsp,
\end{equation}
where $\kernelQ$ is a Markov kernel on
$ \rset^{2d} \times \mcb{\rset^d}$. At a rate $\ps{y}{\nabla U(x)}_+$,
the direction is thus picked according to the kernel $Q$. This
  direction change can be understood as a replacement to rejections
  present in reversible chains and update the direction $y$ in such a
  way that the dynamics eventually targets the distribution
  $\tpi$. Hence $Q$ will be referred to as the \textit{repel
    kernel}. At a rate $\lambdab$, the direction $y$ is simply
\emph{refreshed} by a direct pick from its marginal distribution. This
type of processes can be shown to be ergodic given $\lambdab>0$
following the proof from \cite[Theorem
1]{bouchard:vollmer:doucet:2016} or \cite[Lemma 5.2]{MonmarcheRTP}.

In most work, $Q$ was simply chosen as a Markov kernel on
$(\rset^d \times \msy) \times \mcb{\msy}$, which then defines a PDMP
on $\rset^d \times \msy$ started from $Y_0 \in \msy$. However,
defining a PDMP on $\rset^d \times \rset^d$ is straightforward. It is
just needed to specify $Q((x,y),\cdot)$ for $x \in \rset^d$ and
$y \not \in \msy$ such that $Q((x,y), \msy) = 1$. A particular choice
would be for example to set for $x \in \rset^d$ and $y \not \in \msy$
$Q((x,y),\cdot) = \updelta_{y_0}(\cdot)$ for a fixed $y_0 \in \msy$.

\section{Derivation of new PDMPs for MCMC applications: Forward Event-Chain Monte Carlo}
 \label{sec:derivation-new-pdmps}

 We now propose and investigate new choices for the Markov kernel
 $\kernelQ$ that leaves $\tpi$, defined in \eqref{eq:extended_target},
 invariant for $(P_t)_{t \geq 0}$. In this section, we make an
 informal derivation of such new proposal kernels for the direction at
 event time $(S_n)_{n \in \nset}$ and give some intuitions. A rigorous
 treatment can be found in the supplementary document
 \Cref{sec:form-deriv-new}.

 Our starting point is the characterization of stationarity relying on
 the infinitesimal generator $\generator$ of PDMP processes. This
 generator encodes the infinitesimal changes in time of the semi-group
 $(P_t)_{t\geq 0}$, seen as an operator on a well-chosen class of
 function $\mathscr{C}$, \ie~for any $f \in \mathscr{C}$,
\begin{equation}
  \label{eq:2}
  \generator f= \lim_{t \downarrow 0} \{P_tf-f\}/t \eqsp.
\end{equation}
Here, for $(x,y) \in \rset^{2d}$ and the choice of characteristics
$(\varphi,\kernel,\lambda)$ given by
\eqref{eq:differential_flow}-\eqref{eq:definition_rate} and
\eqref{eq:definition_kernelKQ} (setting $\lambdab=0$ for simplicity),
\begin{multline}
  \label{eq:definition_generator_rough}
  \generator f(x,y) = \ps{y}{\nabla_x f(x,y)} + \ps{y}{\nabla U(x)}_+ \defEns{\int_{\rset^d}f(x,\tilde{y}) Q((x,y),\rmd \ty) - f(x,y) }  \eqsp,
\end{multline}
where $\nabla_x f(x,y)$ is the gradient of the function $x'\mapsto f(x',y)$ at $x$.

If $\tpi$ is a stationary probability measure for $(P_t)_{t \geq 0}$,
i.e. $\tpi P_t = \tpi$ for any $t\geq 0$, then we obtain the condition
$\int_{\rset^{2d}} \generator f \rmd \tpi = 0$ by taking the integral
with respect to $\tpi$ in \eqref{eq:2} and interchanging limit and
integral. Conversely, if $\mathscr{C}$ is sufficiently exhaustive,
this condition is sufficient to show that $\tpi$ is invariant for
$(P_t)_{t \geq 0}$ and is equivalent to the condition
\begin{equation}
\label{eq:extended_global_balance_main}
  \int_{\rset^d} \int_{\rset^d} \ps{y}{\nabla U(x)}_+ f(x,\ty) \kernelQ((x,y),\rmd \ty)  \rmd \loiy(y) =  \int_{\rset^d} \ps{\ty}{\nabla U(x)}_- f(x,\ty)  \rmd \loiy(\ty) \eqsp.
\end{equation}
Note that this relation illustrates the fact that extending the
probability distribution transforms the global-balance condition
\eqref{eq:GB} into an extended global balance, where the update of
$y$ through the repel kernel $\kernelQ$ plays a crucial role.

The previously implemented choices of $\kernelQ$ (\cite{bernard:2009},
\cite{michel:kapfer:krauth:2014}, \cite{bouchard:vollmer:doucet:2016},
\cite{bierkens:fearnhead:roberts:2016}) consist in deterministic
kernels defined for all $(x,y) \in \rset^{2d}$ by
$\msa \mapsto \updelta_{\phi(x,y)} (\msa), \text{ with } \phi:
\rset^{2d} \to \rset^{d}$,
which cancel the integrands and then achieve the \emph{local} balance,
\begin{equation}
\label{eq:sym_local_local_balance}
  \int_{\rset^d} \ps{y}{\nabla U(x)}_+ f(x,\phi(x,y)) \rmd \loiy(y)  =   \int_{\rset^d}\ps{\ty}{\nabla U(x)}_- f(x, \ty) \rmd \loiy(\ty) \eqsp,
\end{equation}
which implies in particular \eqref{eq:extended_global_balance_main}.
As $\loiy$ is rotation invariant, examples of such choices are the
Straight kernel and the Reflection kernel, associated with the
functions $\phi : \rset^{2d} \to \rset^d$ given respectively by
\begin{equation}
\label{eq:bouncy_kernel}
\phi_S : (x,y) \mapsto  -y \eqsp, \quad \phi_R : (x,y) \mapsto (\IdM-2 \rmna{x} \rmna{x}^{\transpose}) y \eqsp,
\end{equation}
where for all $\tx \in \rset^d$,
$ \rmna{\tx} = \nabla U(\tx) / \norm{\nabla U(\tx)} \text{ if } \nabla
U(\tx) \not = 0 $
and $\rmna{\tx} =0$ otherwise.  Both the Straight and Reflection
kernels have been shown to produce speed-ups according to
state-of-the-art methods
\citep{bernard:2009,Michel_20152,bouchard:vollmer:doucet:2016}. However,
they still obey to a stronger-than-necessary local balance
\eqref{eq:sym_local_local_balance}, whereas the original motivation
for PDMP methods is to actually reduce random-walk behavior by
breaking the local detailed balance of the traditional
Hastings-Metropolis methods. Moreover, an additional refreshment step,
\ie~$\lambdab \neq 0$, is needed to ensure ergodicity. We propose now
several repel kernels $\Qkernel$ satisfying the property
\eqref{eq:extended_global_balance_main}, while still being
global. They do not rely on some additionally introduced symmetry but
exploit directly the key role played by the projection along
$\nabla U$ in \eqref{eq:extended_global_balance_main}.

\begin{figure}
\includegraphics[width=1.0\textwidth]{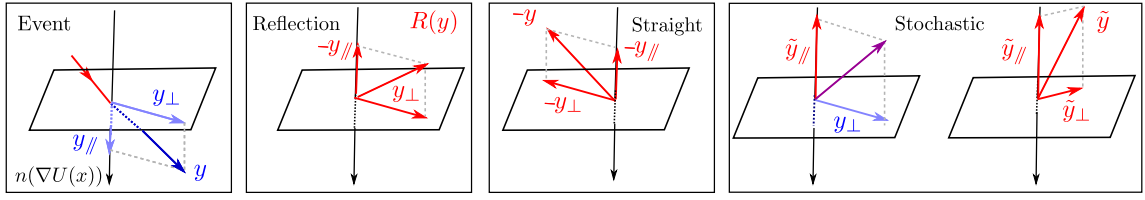}
 \caption{ \footnotesize After an event, a new direction can be picked in a
   deterministic way (Reflection or Straight kernel) or, by exploiting
   the global symmetry around $\nabla U$, the new direction
   $\tilde{y}$ is picked randomly according to the decomposition of
   the kernel $Q$ into $\Kp$ and $\Kn$. In the leftmost figure, $n(\nabla U(x))$ stands for the normalized vector $ \nabla U(x) / \norm{\nabla U(x)}$.}
\label{fig:NewPick}
\end{figure}

Following this remark, the repel kernel $Q$ effect is decomposed into
two contributions: first the update of the direction component along
the gradient $\nabla U$ and second the update of the orthogonal
components. Informally, the Markov kernel $Q$ can be written as the
composition of two Markov kernel $\Kp$ and $\Kn$, respectively on
$\rset \times \mcb{\rset}$ and
$\spana(\nabla U(x))^{\perp} \times \mcbb(\spana(\nabla
U(x))^{\perp})$
and for any $x \in \rset^d$ and $\yp\in \rset$. At the event $S_{n+1}$ and
given the value of the process
$(X_{S_{n+1}}, Y_{S_n}) = (X_{S_n} + (S_{n+1}-S_n)Y_{S_n},Y_{S_n})$,
the new direction is then chosen as follows:
\begin{enumerate}[wide, labelwidth=!, labelindent=0pt, label=(\arabic*)]
\item decompose $Y_{S_n} = Y_{S_{n}}^{\sslash} \nabla U(X_{S_{n+1}}) + Y_{S_n}^{\perp}$ where $Y_{S_n}^{\perp} \in \spana(\nabla U(x))^{\perp}$;
\item first sample $Y_{S_{n+1}}^{\sslash} \sim \Kpp{X_{S_{n+1}}}(-Y_{S_n}^{\sslash},\cdot)$ and second $Y_{S_{n+1}}^{\perp} \sim \Knn{X_{S_{n+1}}}{Y_{S_{n+1}}^{\sslash}}(Y_{S_n},\cdot)$;
  \item  set $Y_{S_{n+1}} = Y_{S_{n+1}}^{\sslash} \nabla U(X_{S_{n+1}}) + Y_{S_{n+1}}^{\perp}$. 
\end{enumerate}
This decomposition for the repel kernel is illustrated on
\Cref{fig:NewPick} and the related pseudo-code simulating a PDMP based
on such a choice for $Q$ is given in \Cref{algo:Forward_MCMC}.

\begin{algorithm}
 \caption{ \label{algo:Forward_MCMC} Forward EC }
  \KwData{Markov kernels $\{(\Kn,\Kp) \, : \, x \in \rset^d, \, \yp \in \rset \}$, refreshment rate $\blambda \geq 0$  and initial points $(X_0,Y_0)$}
 \KwResult{Generic PDMC  $(X_t,Y_t)_{t \geq 0}$ based on \Cref{theo:invariance_BPS_generalized}}
 Initialize $S_0 =0$ and a sequence of \iid~exponential random variables $(E_i^{\text{Ev}},E_i^{\text{Ref}})_{i \geq 1}$ with parameter $1$\\
\mbox{Set  $T_1^{\text{Ref}} = E^{\text{Ref}}_1/$ $\blambda$}\hfill\mbox{\footnotesize\emph{Time before refreshment}}\\
 \For{$n \geq 0$}{  
   \mbox{Set $T_{n+1}^{\text{Ev}} = \inf\{ t \geq 0 \, : \, \int_{0}^t \ps{Y_{S_n}}{\nabla U(X_{S_n} + u Y_{S_n})}_+ \rmd u \geq E^{\text{Ev}}_{n+1}\}$ \shortstack{\footnotesize\emph{Time before}\\\footnotesize\emph{event}}}\\
   \mbox{Set $T_{n+1} = \min(T_{n+1}^{\text{Ev}},T_{n+1}^{\text{Ref}})$ and   $S_{n+1} = S_n + T_{n+1}$}\hfill\mbox{\footnotesize\emph{Time before direction change}}\\
     \mbox{Set   $Y_t = Y_{S_n}$,  $X_{t} = X_{S_n} + (t-S_n) Y_{S_n}$ for $t \in \ooint{S_n,S_{n+1}}$}\hfill\mbox{\footnotesize\emph{Update the position}}\\
 \mbox{\textcolor{white}{Set}  $X_{S_{n+1}} = X_{S_n} + T_{n+1} Y_{S_n}$}\\
 \eIf{$T_{n+1} = T_{n+1}^{\text{Ev}}$}{
   \mbox{$Y_{S_{n}}^{\sslash} = \ps{Y_{S_n}}{\nabla U(X_{S_{n+1}})}$} \hfill   \mbox{\footnotesize\emph{decompose the direction}} \\
   \mbox{$ Y_{S_n}^{\perp} = \{\Id - \nabla U(X_{S_{n+1}})\nabla U(X_{S_{n+1}})^{\transpose}\} Y_{S_n}$} \hfill  \mbox{\footnotesize\emph{along $\nabla U(X_{S_{n+1}})$}}\\
   \mbox{ $Y_{S_{n+1}}^{\sslash} \sim \Kpp{X_{S_{n+1}}}(-Y_{S_n}^{\sslash},\cdot)$}\\
   \mbox{ $Y_{S_{n+1}}^{\perp} \sim \Knn{X_{S_{n+1}}}{Y_{S_{n+1}}^{\sslash}}(Y_{S_n},\cdot)$}\\
\mbox{  set $Y_{S_{n+1}} = Y_{S_{n+1}}^{\sslash} \nabla U(X_{S_{n+1}}) + Y_{S_{n+1}}^{\perp}$ }\\
         \mbox{Set   $T_{n+2}^{\text{Ref}} = T_{n+1}^{\text{Ref}} - T_{n+1}$}\hfill\mbox{\footnotesize\emph{Update the time before refreshment}} \\
        }{
          \mbox{Set $Y_{S_{n+1}} \sim \loiy$}\hfill\mbox{\footnotesize\emph{Refresh the direction}}\\
          \mbox{Set  $T_{n+2}^{\text{Ref}} = E^{\text{Ref}}_{n+2}/$ $\blambda$}\hfill\mbox{\footnotesize\emph{Update the time before refreshment}}
}
 }
\end{algorithm}

Now, the invariance of the extended target distribution $\tpi$ for
$(P_t)_{t\geq 0}$ enforced by \eqref{eq:extended_global_balance_main}
implies simple conditions on the families $\{\Kp\,: \, x \in \rset^d\}$ and $\{\Kn\, : \, x \in \rset^d, \yp \in \rset\}$. Consider the following
two real distributions: $\loiyp$ the distribution of the first
component of $Y$ if $Y \sim \loiy$ and $ \loiym$ the distribution with
density $\yp \mapsto \parentheseLigne{\yp}_-$ with respect to
$\loiyp$. The first condition we need to impose is that for any
$x \in \rset^d$,
\begin{equation}
  \label{eq:3}
  \text{the distribution $\loiym$ is invariant for $\Kp$} \eqsp.
  \tag{\textbf{C$1$}}
\end{equation}
The second condition leads to consider for any $x \in \rset^d$, the conditional distribution
$\tilde{T}^{x,\perp}(\yp,\cdot)$ of the orthogonal projection of
$Y \sim \loiy$ on $\spana(\nabla U(x))^{\perp}$, i.e. the distribution of 
$\{\Id-\nabla U(x) \nabla U(x)^{\transpose}\}Y$, given the component
of $Y$ along $\nabla U(x)$ is equal to $\yp$. The condition
imposes then that for any $x \in \rset^d$ and $\yp \in \rset$,
\begin{equation}
  \label{eq:4}
\text{  $\tilde{T}^{x, \perp}(\yp,\cdot)$ is invariant for the Markov kernel $\Kn$} \eqsp. 
\tag{\textbf{C}$2$}
\end{equation}

In summary, the condition \eqref{eq:4} codes for the fact that the
components on $\spana(\nabla U(x))$ did not trigger any event, so that
their conditional distribution is still
$\tilde{T}^{x,\perp}(\yp,\cdot)$. On the contrary, the direction
component $\yp$ is no longer distributed according to $\loiyp$ but
according to the \emph{reflected-event} distribution $\rho$ defined in
\eqref{eq:3}. This result is formally stated in
\Cref{theo:invariance_BPS_generalized} in the supplementary document.

In the case $\Kp$ and $\Kn$ are the identity kernels,
\ie~$\Kp(\yp,\cdot) = \updelta_{\yp}(\cdot)$ and
$\Kn(\yn,\cdot) = \updelta_{\yn}(\cdot)$ for any $x,\yp,\yn$, then the
PDMC obtained  from \Cref{algo:Forward_MCMC}  recovers the BPS.  But
\Cref{theo:invariance_BPS_generalized} implies many other possible
choices for $\Kn$ and $\Kp$,
 and
therefore lead to a continuum of PDMC methods, forming the Forward
event-chain Monte Carlo class. In the next subsections, we present
  possible choices for $\Kp$ and $\Kn$ and motivate heuristically
  their efficiency. The key idea is to keep the need for refreshment
  and \emph{artificial} noise to a minimum, by ensuring a maximal
  exploration through the new direction picks at events, as they are
  already necessary for correctness.

\subsection{Choices of $\Kp$}
\label{sec:choice_para}
A natural choice for $\Kp$, for $x \in \rset^d$, is to simply choose
the probability measure $\loiym$, if this latter can be efficiently
sampled. In that case, we refer to the resulting scheme as a
\textit{direct sampling} method. If $\loiy$ is the uniform distribution over
the $d$-dimensional sphere $\mathbb{S}^d$, $d \geq 2$, then using spherical
coordinates, we have for all $\msa \in \mcb{\rset}$,
\begin{equation}
  \label{eq:def_loiy_para_sphere}
  \loiym(\msa) = \int_{0}^{\uppi/2} \1_{\msa}(-\cos(\theta)) \, \frac{\cos(\theta) \sin^{d-2}(\theta)}{d-1} \rmd \theta = \int_{-1}^0 \1_{\msa}(v) \frac{(-v) (1-v^{2})^{(d-3)/2}}{d-1} \rmd v \eqsp.
\end{equation}
Therefore, $ \loiym$ can be efficiently sampled since if $V$ is a
uniform random variable on $\ccint{0,1}$, then it is straightforward
to verify that $(1-V^{2/(d-1)})^{1/2}$ has distribution $\loiym$. In
the case where $\loiy$ is the $d$-dimensional standard Gaussian
distribution, then it is easy to check that
$\msa \mapsto \loiym(-\msa)$ is the $\chi$-distribution with $2$
degrees of freedom, as proposed by
\cite{vanetti:et:al:2017} building on earlier
version of this work,  with $\Kn = \Id$.

It is also possible to set $\Kp$ to be a Markov kernel defined by a
Metropolis-Hastings algorithm designed to sample from $\loiym$. In
that case, we refer to the resulting scheme as a \textit{Metropolis
  sampling} method.  One such example would be a random walk or
independent Metropolis-Hastings algorithm on $\ccint{-1,0}$ with
Gaussian or uniform noise. Explicit expressions for the associated
Markov kernels for $\msy = \sphere^{d-1}$ are given in the
supplementary document \Cref{sec:choices-kp}.  The choices for $\Kp$
are naturally not bounded to these schemes. For instance, it is
possible to define a mixture of kernels, as e.g. a direct-sampling
kernel with an identity one.

\subsection{Choices of $\Kn$}
\label{sec:choice_ortho}
A trivial choice is $\Kn = \Id$, which is the case of the BPS and
standard EC processes, but which both rely on a refreshment step.  It
can be advantageous to set $\Kn$ differently in order to improve the
exploration of the state space and to therefore insure the
  ergodicity of the process, while aiming at setting the refreshment rate $\lambdab$
in \eqref{eq:definition_rate} to zero.
We propose next several possibilities for $\Kn$ for the specific case $\msy = \sphere^{d-1}$.


The idea is to rely on the randomization achieved on the parallel
component by $\Kp$ to minimize at most the randomization on the
orthogonal components. As, for any $x \in \rset^d$, $\yp \in \ccint{0,1}$,
$\tilde{T}^{x,\perp}(\yp,\cdot)$ \footnote{In the case where $\loiy$
  is the uniform distribution on $\sphere^{d-1}$,
  $\tilde{T}^{x,\perp}(\yp,\cdot)$ defined in \eqref{eq:def_tilde_T}
  is the uniform distribution on the $d-1$-sphere of
  $\spana(\nabla U(x))^{\perp}$ with radius $(1-\yp^2)^{1/2}$, for all
  $x \in \rset^d$ and $\yp \in \ccint{-1,1}$. In the case where
  $\loiy$ is the $d$-dimensional standard Gaussian distribution,
  $\tilde{T}^{x,\perp}(\yp,\cdot)$ is the $(d-1)$-dimensional standard
  Gaussian, for all $x \in \rset^d$ and $\yp \in \rset$.} is rotation invariant, we
characterize the various choices for $\Kn$ by considering for any
$x \in \rset^d$ a probability distribution $\nu^x$ on the set of
orthogonal transformations $\matOrt^x$ on
$\spana(\nabla U(x))^{\perp}$ and define the Markov kernels for any
$\yn \in \spana(\nabla U(x))^{\perp} \setminus \{0\} $ and
$\msa \in \spana(\nabla U(x))^{\perp}$ by
\begin{align}
\label{eq:def_Kn_1_p_ortho}
  \Kn_{\text{naive}}(\yn,\msa) &= \int_{\matOrt^x} \1_{\msa}\parenthese{(1-\yp^2)^{1/2} \mathrm{O}(\yn/\norm{\yn})} \rmd \nu^x(\mathrm{O}) \eqsp, \\
  \label{eq:def_Kn_2_p_ortho}
   \Kn_{\text{pos}}(\yn,\msa) &= \int_{\matOrt^x} \1_{\msa}\parenthese{(1-\yp^2)^{1/2}\sign(\ps{\yn}{\rmO \yn}) \mathrm{O}(\yn/\norm{\yn})} \rmd \nu^x(\mathrm{O}) \eqsp.
\end{align}
In the case $\yn = 0$, just choose a deterministic point in the $d-1$-dimensional sphere of $\spana(\nabla U(x))$. 
Both kernels then admit $\tilde{T}^{x,\perp}(\yp,\cdot)$ as invariant
distribution for any $x \in \rset^d$, $\yp \in \ccint{0,1}$ when $\loiy$ is the uniform distribution on $\sphere^{d-1}$.  Indeed, the
result for $\Kn_{\text{naive}}$ is straightforward. As for
$\Kn_{\text{pos}}$, using that $\tilde{T}^{x,\perp}(\yp,\cdot)$ is
rotation invariant, we get easily that
$\tilde{T}^{x,\perp}(\yp,\cdot) \Kn_{\text{pos}}$ is rotation
invariant with support included in the $(d-1)$-dimensional sphere of
$\spana(\nabla U(x))^{\perp}$ with radius $(1-\yp^2)^{1/2}$, therefore
$\tilde{T}^{x,\perp}(\yp,\cdot) \Kn_{\text{pos}} =
\tilde{T}^{x,\perp}(\yp,\cdot) $.
Sampling from $\Kn_{{\text{naive}}}(\yn,\cdot)$ (resp.
$\Kn_{\text{pos}}(\yn,\cdot)$) comes down to sampling $O$ from $\nu^x$
and set $\Yn = O(\yn)$
(resp. $\Yn=\sign(\ps{\yn}{O(\yn)})O(\yn)$). Contrary to
$\Kn_{{\text{naive}}}$, using $\Kn_{\text{pos}}$ imposes that the new
direction $\Yn$ satisfies $\ps{\yn}{\Yn} \geq 0$ and therefore avoids
that the position backtracks. Finally, this description can be further
generalized in the case of $\loiy$ being the standard Gaussian
distribution by adding an additional norm sampling step.

This class of Markov kernels $\Kn$ offers a great freedom on the
randomness we want to use in the algorithm and potentially avoid
random-walk behaviour by changing a full and global refreshment into a
sparse and orthogonal one. Indeed, if we choose for $\nu^x$, the
uniform distribution on $\matOrt^x$, then the noise produced by the
method is significant. In fact, when $\msy=\sphere^{d-1}$, $\Kn$ is
equal to $T^{x,\perp}(\yp,\cdot)$ for any $x \in \rset^d$ and
$\yp \in \ccint{-1,1}$. The resulting scheme is referred to as a
\textit{full-orthogonal refresh} method. In the case where $\loiy$ is
the $d$-dimensional standard Gaussian distribution, this choice of
$\Kn$ can be extended by a norm resampling to recover
$T^{x,\perp}(\yp,\cdot)$. It was also proposed in
\cite{wu:robert:2017}, after earlier versions of this work, in the
case where $\Kp = \Id$. However such a choice, while ensuring
ergodicity \citep{wu:robert:2017}, leads to a quasi-refreshment at
every event and introduces strong noise and random-walk behavior, see
e.g. Figure~\ref{fig:Gauss_Num_Ref_SparseFull} in supplement. The
noise can be reduced by considering a mixture with the identity
kernel, but this asks for a fine tuning and is similar to choosing
$\Kn = \Id$ and $\lambdab\neq 0$, see
Figures~\ref{fig:Gauss_Num_IntRefs} and
\ref{fig:Gauss_Num_IntGaussRef}.

On the contrary, if we explore another strategy based on a kernel
$\Kn$ allowing only for a partial refreshment and choose for $\nu^x$
to be a probability measure such that its support is contained in the
subspace of orthogonal matrices, $\matOrt^p_x$, which only act on
$p$-dimensional space, $p \in \{2,\ldots,d-1\}$, \ie
\begin{equation*}
  \matOrt^p_x = \ensemble{\mathrm{O} \in \matOrt_x}{\ker(\Id-\rmO) = d-1-p} \eqsp,
\end{equation*}
then the noise can be considerably smaller taking for example $p =2$
and $d$ large. In addition, distributions on $\matOrt^p_x$ can be very
cheap to compute for small $p$ as we will see. In the case where $\nu^x$ is the uniform distribution on $\matOrt^p_x$, the choice of $\Kn=\Kn_{{\text{naive}}}$ and $\Kn=\Kn_{\text{pos}}$ defined by \eqref{eq:def_Kn_1_p_ortho} and \eqref{eq:def_Kn_2_p_ortho} respectively,  lead to scheme referred in the following as naive or positive \textit{$p$-orthogonal refresh}.

From a practical perspective, distributions $\nu^x$ on $\matOrt_x$ can
be easily derived from a probability distribution on the set of
orthogonal matrices $\matOrt(d)$, with its Borel $\sigma$-field
$\mcbb(\matOrt(d))$. Indeed, it suffices to compute an orthogonal
basis for $\spana(\nabla U(x))^{\perp}$ which can be done using the
Gram-Schmidt process on the canonical basis
$(\bfe_i)_{i \in \{1,\ldots,d\}}$. An other solution, which is computationally cheaper, is to find $i \in \{1,\ldots,d\}$ such that $\bfe_i \not \in \spana(\nabla U(x))$, and set
\begin{equation*}
  \tilde{\bfe}_j(x) = \defEns{\Id -2(\nabla U(x) - \bfe_i)(\nabla U(x) - \bfe_i)^{\transpose}}\bfe_j \eqsp, \, \text{ for $j \in \{1,\ldots,d\} \setminus \{i\}$} \eqsp.
\end{equation*}
Then, we can easily check, since
$\Id -2(\nabla U(x) - \bfe_i)(\nabla U(x) - \bfe_i)^{\transpose}$ is
an orthogonal matrix, that
$( \tilde{\bfe}_j(x))_{j \in \{1,\ldots,d\} \setminus \{i\}}$ is an
orthonormal basis of $\spana(\nabla U(x))^{\perp}$. We can observe
that this computation has complexity $O(d^2)$ which can be prohibitive
and that is why we propose different constructions of probability measures $\nu^x$ without this constraint.

For example, we consider in \Cref{sec:numexp}, the case where, for
$x\in \rset^d$, $\nu^x$ is the distribution of the random variable $O$
defined as follows. Consider two $d$-dimensional Gaussian random
variables $G_1$ and $G_2$, and the two orthogonal vectors $e_1$ and
$e_2$ in $\spana(\nabla U(x))^{\perp}$ defined by the Gram-Schmidt
process and based on
$\tilde{G}_1= (\Id -\nabla U(x) \nabla U(x)^{\transpose}) G_1$ and
$ \tilde{G}_2 = (\Id -\nabla U(x) \nabla U(x)^{\transpose}) G_2$, \ie
\begin{equation*}
  e_1= \tilde{G}_1/ \normLigne{\tilde{G}_1} \eqsp, \, \,  e_2 = (\tilde{G_2} - \psLigne{e_1}{ \tilde{G_2} }e_1)/\normLigne{\tilde{G_2} - \psLigne{e_1}{ \tilde{G_2} }e_1} \eqsp.
\end{equation*}
Then, the random orthogonal transformation $O$ is defined by
\begin{equation}
  \label{eq:gram_schmidt_ran}
  O_{\theta} = \{\cos(\theta) e_1 + \sin(\theta) e_2\}  e_1^{\transpose} + \{ \sin(\theta) e_1 - \cos(\theta) e_2\}e_2^{\transpose} \eqsp,
\end{equation}
for $\theta \in \ccint{0,2\uppi}$, belongs to $\matOrt^2_x$ almost
surely.  The choice of $\theta$ naturally impacts the randomization and the case $\theta=\uppi/2$ will be referred to as the \emph{orthogonal switch}  refresh and the case where the parameter $\theta$ can be itself   random \emph{ran-p-orthogonal} refresh. 

In the case where $\loiy$ is the $d$-dimensional
standard Gaussian distribution, we can consider the auto-regressive
kernel on $\spana(\nabla U(x))^{\perp}$, defined for any
$\yn \in \spana(\nabla U(x))^{\perp}$ and
$\msa \in \mcbb(\spana(\nabla U(x))^{\perp})$ by
\begin{equation*}
  \Kn(\yn,\msa) = (2\uppi)^{-d/2}\int_{\rset^d} \1_{\msa} \parenthese{\rho \yn + \sqrt{1-\rho^2} (\Id - \nabla U(x) \nabla U(x)^{\transpose})\ty} \rme^{-\norm{\ty}^2/2} \rmd \ty \eqsp,
\end{equation*}
where $\rho \in \ccint{0,1}$.  In other word, starting from $\yn$, the
component along $\spana(\nabla U(x))^{\perp}$ of the new direction is
set to be
$\rho \yn + (1-\rho^2) (\Id - \nabla U(x) \nabla
U(x)^{\transpose})Y_1$, where $Y_1$ is a $d$-dimensional standard
Gaussian random variable. The resulting PDMP-MCMC with $\Kp = \Id$ or
$\Kp = \loiym$ was proposed by \cite{vanetti:et:al:2017}.

As for $\Kp$, $\Kn$ can be a mixture of the identity kernel and a
partial refreshment of the orthogonal components: $p_r \Id + (1-p_r) \tKn$
such that $p_r \in \ccint{0,1}$ and $\tilde{T}^{x,\perp}$ is invariant for
$\tKn$ for any $x \in \rset^d$. This step corresponds to a transformation of the
sampling $Y_{n+1}^{\perp}$ in \Cref{algo:Forward_MCMC} into
$Y^{\perp}_{n+1} = B Y_{S_{n}} +(1-B)\tilde{Y}_{n+1}^{\perp}$, where
$B$ is a Bernoulli random variable with parameter $p_r$ and
$\tilde{Y}_{n+1}^{\perp} \sim \tKnn{X_{S_{n+1}}}(\projnn{X_{S_{n+1}}}(Y_{S_{n}}),\cdot) $.

\subsection{About  refreshment strategy}
\label{sec:refresh_strat}
Choices of $\Kn$ different from the identity corresponds to a partial
refresh of the orthogonal components of the direction. However, it can be
advantageous to not use this kind of refreshment at any events.

As proposed above, a first option is to choose a mixture with the
identity kernel and to choose the parameter $p_r$ accordingly. It can be
interesting to control the partial refreshment through the time
parameter directly, as fixing a \emph{refreshment} time to $\rmT$ can
simplify implementation. A second option is thus as follows. First, we
extend the state space $\mse = \rset^d\times \msv$ to $\mse \times
\{0,1\}$ and consider two Markov kernels $Q_0,Q_1$ on $\rset^{2d}
\times \mcb{\rset^d}$ associated with $\Kn_0,\Kn_1,\Kp_0,\Kp_1$
satisfying the conditions of
\Cref{theo:invariance_BPS_generalized}. We now consider the PDMP
$(\bX_t,\bY_t,\bB_t)_{t \geq 0}$ corresponding to the differential
flow $\bvarphi_t(x,y,b) = (x+ty,y,b)$ for any $(x,y,b) \in \mse \times
\{0,1\}$, event rate $\lambda$ given by \eqref{eq:definition_rate}
with $\lambdab=0$ and Markov kernel $\bQ$ on $(\mse \times \{0,1\} )
\times \mcb{\msy \times \{0,1\}}$ defined for any $(x,y,b) \in \mse
\times \{0,1\}$, $\msa \in \mcb{\msy}$ by
\begin{equation*}
  \bQ((x,y,b),\msa\times \{1\}) = Q_b((x,y),\msa) \eqsp. 
\end{equation*}
If $Q_0$ differs from the identity kernel, $Q_1$ is the identity
kernel and the extra variable $(\bB_t)_{t\geq 0}$ is updated to $0$
every time $\rmT>0$. This method produces a partial refreshment
through $Q_0$ at each event following directly an update of $b$. More
details and a pseudocode for this PDMP can be found in
\Cref{sec:refresh_strat_app} in the supplement.  In particular,
\Cref{fig:Gauss_Num_FixMix} shows that the same decorrelation is
obtained from both strategies.  

It is also possible to transform the stochastic refreshment step
  ruled by the Poisson process of rate $\lambdab$ by a refreshment
  process at every time $\rmT$ by considering a collection of PDMP of
  length $\rmT$ instead of a single PDMP. A pseudocode is given in
  \Cref{algo:chain_length} in the supplement. 

\subsection{PDMC and Potential Factorization}

When the potential $U$ can be written as a sum of terms, $ U(x) =
\sum_{i=1}^d U_i(x)$ or considering directly the decomposition of the
gradient $\nabla U$ over the direction, it can be convenient to
exploit this decomposition through the implementation of the
factorized Metropolis filter \citep{michel:kapfer:krauth:2014}, for
example to exploit some symmetries of the problem or reduce the complexity 
\citep{michel:2017}. It finds its equivalent in PDMC by
considering a superposition of Poisson processes
\citep{peters:dewith:2012,michel:kapfer:krauth:2014,bouchard:vollmer:doucet:2016}. The results developed in \Cref{sec:derivation-new-pdmps} can be
generalized using this property, as we explain in more details in the
supplement \Cref{sec:DPDMC}.

\section{Numerical Experiments}
\label{sec:numexp}

We restrict our numerical studies to the case $\msv = \sphere^{d-1}$
and $\loiy$ is the uniform distribution on $\sphere^{d-1}$. After
specifying the comparison methods in \Cref{sec:Num_Compar}, we consider
four different types of target density $\pi$: ill-conditioned Gaussian
distributions in \Cref{sec:gaussian}, a Poisson-Gaussian Markov random
field in \Cref{sec:poisson-gauss}, mixtures of Gaussian distributions
in \Cref{sec:mixgaussian} and finally a posteriori distributions
coming from logistic regression problems in \Cref{sec:logreg}.

Codes used for these numerical experiments are available at {\small \url{https://bitbucket.org/MNMichel/forwardec/src/master/}}.

\subsection{Comparing schemes}
\label{sec:Num_Compar}

Similarly to \cite{bernard:2009}, \cite{bouchard:vollmer:doucet:2016}
and \cite{bierkens:fearnhead:roberts:2016}, for a fixed test function
$h$ and PDMC $(X_t,Y_t)_{t \in\ccint{0,t_f}}$ for a final time $t_f \geq
0$, we consider the estimator of $\int_{\rset^d} h(x) \pi(x) \rmd x$,
$ \hh_n = n^{-1}\sum_{i=1}^{n} h(X_{\delta i})$,
where $n = \floor{t_f/\delta}$ and $\delta$ is a fixed step-size. To
compare the different schemes, we then consider different
criteria. First, we define the autocorrelation function associated
with $h$ at lag $k \in \nset$ by
$  C_h(k) = (n-k)^{-1}\sum_{i=0}^{n-k-1} \{(h(X_{\delta i})h(X_{\delta (i+k)})-m_h^2)/\sigma^2_h \}$, 
where $m_h$ and $\sigma_h$ are either set to
$\int_{\rset^d} h(x) \pi(x) \rmd x$ and
$\int_{\rset^d} (h(x)-m_h)^2 \pi(x) \rmd x$ when it is possible to
calculate these values or to approximations of these quantities
obtained after a long run. Other
criteria that we investigate amongst the different schemes is their
integrated correlation time, $\tau^{\rmint}_h$ defined by
$  \tau^{\rmint}_h = 1/2 + \sum_{k=1}^{N_{\rmint}} (1-k/N)C_h(k)$ where $ N_{\rmint} = \inf\{ k \in \nset\,: \, \text{ for all } i \geq k, \,  \abs{C_h(i)} \leq 10^{-3} \} $,
 and their Effective Sample Size (ESS) given by $
\ESS^h = N_{\tau_{\rmint}}/(2 \tau^{\rmint}_h ) $. Finally, we
  stress the importance of using different test functions $h$ as PDMC,
  thanks to their ballistic trajectories, can lead to fast
  decorrelation for some functions $h$, while showing a very slow
  decay for others or even lack of ergodicity, see
  \eg~Figure~\ref{fig:Gauss_Num_BPSOrth} in the supplement.

To be able to have a fair comparison in terms of computational
efficiency, we plot the autocorrelations as a function of
the averaged number of events per samples $n_{\delta}$, corresponding
to the averaged number of gradient evaluations per samples. In
practice, it simply leads to the sequence $(C^h(k/n_{\delta}))_{k \in
  n_{\delta} \nset}$. The same procedure is done for the integrated
correlation time, which ends up being multiplied by $n_{\delta}$.
Box plots are based on $100$ runs of $10^5$ samples separated by a
fixed $\delta$ which will be specified.\\

In the following experiments, we compare the performance of the
following schemes:
\begin{itemize}
\item Forward No Ref: direct-sampling scheme with no refreshment of
  the orthogonal components and $\lambdab =0$.  This method
  corresponds to the choice of $\Kp = \rho$ and $\Kn = \Id$ in
  \Cref{algo:Forward_MCMC}. As there is no refreshment,
    particular care on testing ergodicity of the process has to be taken.
\item Forward All Ref: direct-sampling scheme with refreshment at
  every event according to an orthogonal switch and $\lambdab =0$.
  This method corresponds to the choice of $\Kp = \rho$
  and $\Kn = \Kn_{\text{pos}}$ defined by \eqref{eq:def_Kn_2_p_ortho}, where
  $\nu^x$ is the distribution of the random variable defined by
\eqref{eq:gram_schmidt_ran} with $\theta =\uppi/2$.
\item Forward Ref: direct-sampling scheme with refreshment at an event
  every time $\rmT$ according to an orthogonal switch and
  $\lambdab =0$ see \Cref{sec:refresh_strat}.  This method corresponds to the pseudo-code 
  \Cref{algo:update_Ext_Forward_MCMC} in the supplementary document using the two kernels $Q_0$ and
  $Q_1$ associated with $\Kp_0 = \Kp_1=\rho$ and $\Kn_0=  \Kn_{\text{pos}}$  where
$\nu^x$ is the distribution of the random variable defined by
\eqref{eq:gram_schmidt_ran} with $\theta =\uppi/2$.
\item Forward Full Ref: direct-sampling scheme with no refreshment of
  the orthogonal components and a full refreshment of the direction
  every $\rmT$, see \Cref{sec:refresh_strat}. This method corresponds
  to \Cref{algo:chain_length} in the supplementary document and to the choice $\Kp = \rho$, $\Kn=\Id$.
\item BPS Full Ref: reflection scheme with no refreshment of
  the orthogonal components and a full refreshment of the direction
  every $\rmT$,  see \Cref{sec:refresh_strat}. This method  corresponds to \Cref{algo:chain_length} and  to the choice $\Kp = \Id$, $\Kn=\Id$.
\item BPS No Ref: reflection scheme ($\Kp=\Id$) with no refreshment of
  the orthogonal components ($\Kn=\Id$) and no full refreshment
  ($\lambdab=0$). As there is no refreshment, particular care on
    testing ergodicity of the process has to be taken.
\end{itemize}
Out of completeness, we will also display the performance of the
Hamiltonian Monte Carlo and the Zig Zag schemes for the anisotropic
Gaussian experiments. As we are comparing the efficiency of
  given Markov kernels, we did not include schemes based on metric
  adaptations and fits, as NUTS \citep{nuts}.

\subsection{Anisotropic Gaussian distribution}
\label{sec:gaussian}

We consider the problem of sampling from a $d$-dimensional
zero-mean anisotropic Gaussian distribution in which the eigenvalues of the
covariance matrix $\Sigma$ are log-linearly distributed between $1$ and $10^6$,
such as in \cite{hmcdetbal}, \ie~ we set for any $i,j\in\{1,\ldots,d\}$, and $d \geq 2$,
\begin{equation}
  \label{eq:Gaussian}
   \Sigma_{i,j} = \updelta_{i,j} \exp\left(\frac{6(i-1)}{d-1}\log 10\right) \eqsp.
\end{equation}
We develop the calculations of the event times for a Gaussian
distribution in \Cref{sec:sampling-event-times} of the
supplement.

\begin{figure}[h]
\begin{center}
\includegraphics[width=1.0\textwidth]{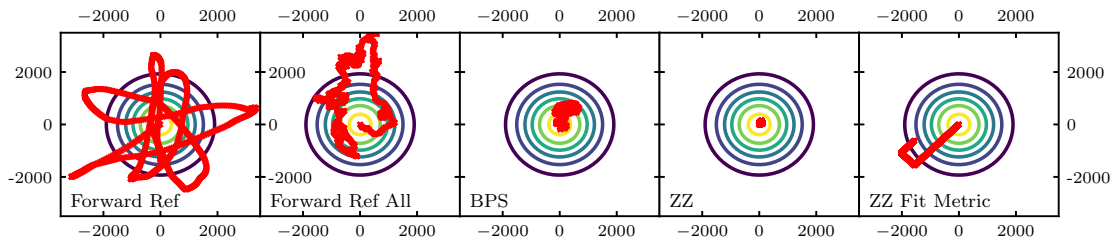}
\caption{ \footnotesize First $1000$ samples (red crosses) for the $400$-dimensional
  zero-mean Gaussian distribution with covariance matrix given by
  (\ref{eq:Gaussian}) (contoured) generated from the same initial
  position. Successive positions are separated by the same number of
  events ($\sim 55$).}
\label{fig:Gauss_Num_Plot}
\end{center}
\end{figure}

We tuned the refreshment rate for Forward Ref and BPS Full Ref in
order to achieve the fastest decorrelation for the potential $U$ at
$d=400$ ($T=500$, corresponding roughly to an average of $55$ events),
as $U$ is not sensitive to the ill-conditioned nature of the
distribution and requires mixing on all dimensions.  To allow for an
easy comparison with BPS, Forward Full Ref refreshment rate is also
set to the same rate. The Hamiltonian Monte Carlo scheme is optimized
through an adaptive implementation in order to achieve an acceptance
rate $\approx 0.6$ \citep{beskos:pillai:roberts:et:al:2013}. The ZZ
algorithm is run according to a random basis of vectors (ZZ) and to
the eigenvectors basis (ZZ Fit Metric).  We also simulated a standard
EC scheme factorized according to the eigenvector basis and refer to
it in the following by Optimized EC, playing the role of an ideal
reference. The difference between the EC and ZZ schemes lies in the
fact that the EC successively updates the position according to each
basis vector successively, whereas the ZZ updates the position
according to all simultaneously. Finally, for this highly-symmetrical
distribution, the schemes without refreshment (BPS No Ref and Forward
No Ref) are not ergodic, as they would stick to a plane. Comparison to
a standard Hastings-Metropolis scheme can be found in the supplement
in \Cref{sec:numer-metro}, showing the limited efficiency of a
standard random walk for this type of distribution.

\begin{figure}[h!]
\begin{center}
\includegraphics[width=1.0\textwidth]{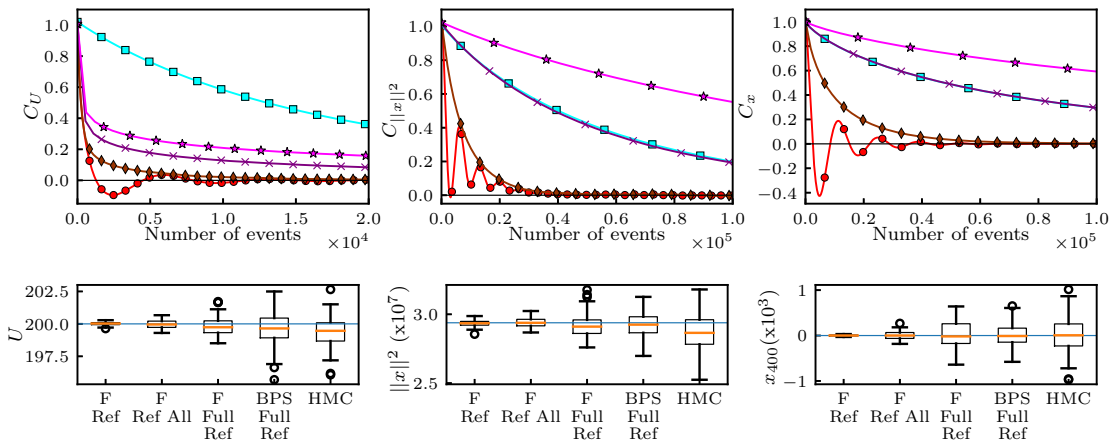}
\caption{ \footnotesize\textbf{Top:} Autocorrelation functions $C$ of
  $U$ (\textbf{Left}), $\norm{x}^2$ (\textbf{Middle}) and $x$
  (\textbf{Right}) for the zero-mean Gaussian distribution with
  covariance matrix given by (\ref{eq:Gaussian}) and $d=400$ for
  Forward Ref (red, circles), Forward Ref All (maroon, thin diamond),
  Forward Full Ref (purple, cross), BPS Full Ref (cyan, square) and
  HMC (magenta, star).\textbf{Bottom:} Boxplots for $U$, $\norm{x}$
  and highest-variance component $x_{400}$. For the box plots, samples
  are separated by an averaged of $55$ events (or gradient evaluations
  for HMC). }
\label{fig:Gauss_Num_Autocorr_Ref}
\end{center}
\end{figure}

\begin{figure}[h!]
\begin{center}
\includegraphics[width=1.0\textwidth]{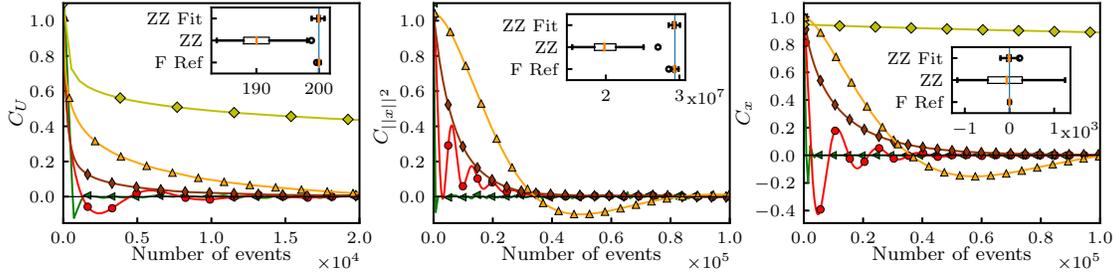}
\caption{ \footnotesize Autocorrelation functions $C$ of $U$
  (\textbf{Left}), $\norm{x}^2$ (\textbf{Middle}) and $x$
  (\textbf{Right}) for the zero-mean Gaussian distribution with
  covariance matrix given by (\ref{eq:Gaussian}) and $d=400$. for
  Forward Ref (red, circles), Forward Ref All (maroon, thin diamond),
  ZZ (light green, diamond), ZZ Fit Metric (yellow, up triangle) and
  Optimized EC (green, right triangle). The autocorrelation function
  of the norm for ZZ is not displayed out of convergence issue. The
  insets are the respective boxplots for $U$, $\norm{x}$ and
  highest-variance component $x_{400}$. For the box plots, samples are
  separated by an averaged of $55$ events. }
\label{fig:Gauss_Num_Autocorr_Straight}
\end{center}
\end{figure}

\begin{figure}[h!]
\begin{center}
\includegraphics[width=1.0\textwidth]{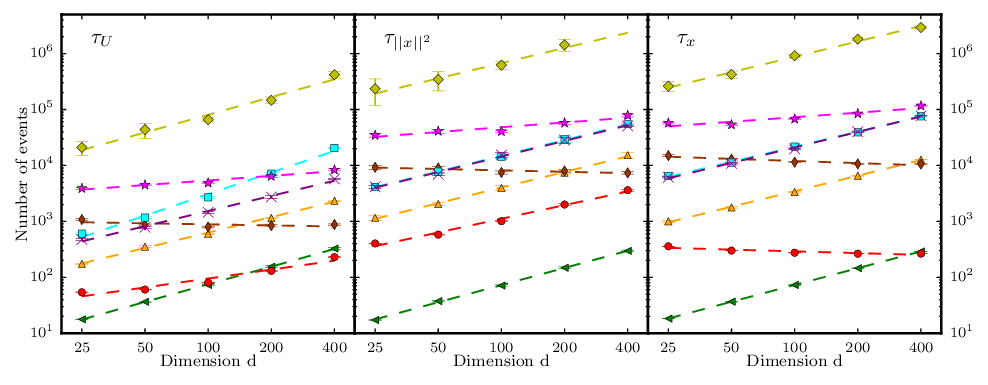}
\includegraphics[width=1.0\textwidth]{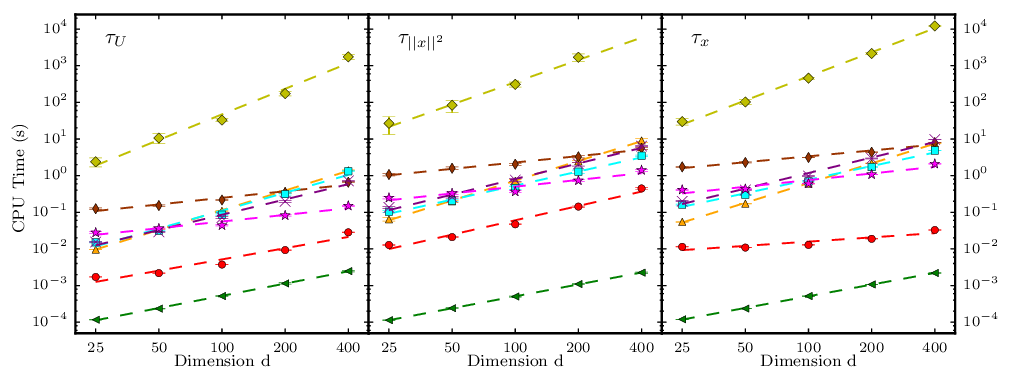}
\caption{ \footnotesize Integrated autocorrelation times $\tau$ of $U$
  (\textbf{Left}), $\norm{x}^2$ (\textbf{Middle}) and $x$
  (\textbf{Right}) for the zero-mean Gaussian distribution with
  covariance matrix given by \eqref{eq:Gaussian} for Forward Ref (red,
  circles), Forward Ref All (maroon, thin diamond), Forward Full Ref
  (purple, cross), BPS Full Ref (cyan, square), HMC (magenta, star) ZZ
  (light green, diamond), ZZ Fit Metric (yellow, up triangle) and
  Optimized EC (green, right triangle), in number of events
  (\textbf{Top}) and amount of CPU time (\textbf{Bottom}). Dashed
  lines stand for the fit $\tau_h = Ad^{z}$. Error bars may be covered
  by the markers.}
\label{fig:Gauss_Num_Int}
\end{center}
\end{figure}

Figure~\ref{fig:Gauss_Num_Plot} exhibits section plots showing the
first $1000$ samples generated from an initial position at the
origin. This qualitative picture is confirmed by the autocorrelation
functions displayed on Figure~\ref{fig:Gauss_Num_Autocorr_Ref} for the
reflection-kernel schemes (HMC included) and on
Figure~\ref{fig:Gauss_Num_Autocorr_Straight} for the straight-kernel
schemes and the scaling of integrated autocorrelation times with the
dimension in \Cref{fig:Gauss_Num_Int}, in terms of events and of CPU
times, to account for extra complexities, as in particular for the
ZZ scheme in the general case. The corresponding fit results are given
in \Cref{tab:Scaling_Gauss}.

In summary, Forward Ref achieves clear quantitative acceleration in
comparison to the other methods, excluding the ideal Optimized EC, and
exhibits antithetic autocorrelations, showing the reduction of any
random-walk behavior. Moreover, as the scaling with the dimension $d$
is smaller than for the other methods excepted HMC and Forward
  All Ref), this acceleration increases with the dimension of the
target distribution. Forward All Ref exhibits the smallest
  scaling, $0$ in terms of number of events and $0.6$ in terms of CPU
  time, matching then HMC. This acceleration is due to both the
direct-sampling $\Kp$ and the sparse orthogonal switch $\Kn$. Forward
Full Ref has the same $\Kp$ than Forward Ref and Forward Ref All but a
different $\Kn$ and exhibits a fast decorrelation of $U$ compared to
BPS, but is overall slower than Forward Ref and Forward Ref All. Even
set to an optimal refreshment time $\rmT$ > 0, BPS is slower and, set
to an identic sparse orthogonal switch $\Kn$, converges even more
slowly, see in the supplement \Cref{sec:numer-comp-diff} and
Figure~\ref{fig:Gauss_Num_BPSOrth}. Finally, regarding tuning
sensitivity, the parameter-free version, Forward Ref All, shows an
even better scaling than Forward Ref while requiring no tuning, which
makes it a competitive option for high dimensions. The refreshment
time $\rmT$ requires indeed no crucial tuning for Forward Ref, on the
contrary of BPS or Forward Full Ref, as illustrated by
Figure~\ref{fig:Gauss_Num_IntForward},
Figure~\ref{fig:Gauss_Num_IntRefs} and
Figure~\ref{fig:Gauss_Num_IntBPS} in supplement, which show that
varying $\rmT$ from $50$ to $1000$ leads up to more than a $30-$fold
increase of the maximal integrated autocorrelation time for BPS and
Forward Full Ref, whereas it is less than a $4-$fold increase for
Forward Ref. The impact of the choice of $p$ and $\theta$ in $\Kn$ are
also investigated in Appendix \ref{sec:rate} and appears not to be
critical.

\begin{table}
  \caption{\small  Scaling $z$ of the integrated autocorrelation times
    $\tau_h$ with the dimension of the anisotropic Gaussian
    distribution in terms of events (\textbf{Top}) or CPU time (\textbf{Bottom}), for the least-squares fit $\tau_h=Ad^z$ by
    Levenberg-Marquardt gradient method.  
    \label{tab:Scaling_Gauss}}
\setlength\tabcolsep{0.1cm}
\small
    \begin{tabular}{p{0.75cm}llllllll}
  \scriptsize Events           &F. Ref           &F. Ref All      &F. Full      &BPS Full       &ZZ             &ZZ Fit         &HMC           &Opt. EC\\\hline
 $U$         &0.53\scriptsize$\pm$0.01  &-0.06\scriptsize$\pm$0.02  &0.90\scriptsize$\pm$0.01  &1.28\scriptsize$\pm$0.02  &1.05\scriptsize$\pm$0.06  &0.92\scriptsize$\pm$0.01  &0.27\scriptsize\scriptsize$\pm$0.02 &1.05\scriptsize$\pm$0.01\\ 
            
$\norm{x}^2$ &0.81\scriptsize$\pm$0.01   &-0.08\scriptsize$\pm$0.03  &0.93\scriptsize$\pm$0.03  &0.93\scriptsize$\pm$0.02  &0.9\scriptsize$\pm$0.3    &0.93\scriptsize$\pm$0.05  &0.28\scriptsize$\pm$0.04 &1.02\scriptsize$\pm$0.01\\
            
      $x$    &-0.10\scriptsize$\pm$0.01 &-0.13\scriptsize$\pm$0.01  &0.92\scriptsize$\pm$0.01  &0.89\scriptsize$\pm$0.01  &0.92\scriptsize$\pm$0.09  &0.92\scriptsize$\pm$0.01  &0.27\scriptsize$\pm$0.02 &1.00\scriptsize$\pm$0.01\\  \hline
 \scriptsize  CPU           &F. Ref           &F. Ref All      &F. Full      &BPS Full       &ZZ             &ZZ Fit         &HMC           &Opt. EC\\\hline
 $U$    &    1.00\scriptsize$\pm$0.01  &0.60\scriptsize$\pm$0.02  &1.39\scriptsize$\pm$0.02  &1.62\scriptsize$\pm$0.02  &2.3\scriptsize$\pm$0.1  &1.78\scriptsize$\pm$0.01  &0.60\scriptsize\scriptsize$\pm$0.02 &1.1\scriptsize$\pm$0.01\\
$\norm{x}^2$ &1.30\scriptsize$\pm$0.01  &0.60\scriptsize$\pm$0.03  &1.41\scriptsize$\pm$0.03  &1.28\scriptsize$\pm$0.02  &2.0\scriptsize$\pm$0.3  &1.79\scriptsize$\pm$0.05  &0.61\scriptsize\scriptsize$\pm$0.04 &1.08\scriptsize$\pm$0.01\\
      $x$    &0.35\scriptsize$\pm$0.01  &0.53\scriptsize$\pm$0.01  &1.40\scriptsize$\pm$0.01  &1.23\scriptsize$\pm$0.01  &2.2\scriptsize$\pm$0.1  &1.79\scriptsize$\pm$0.01  &0.60\scriptsize\scriptsize$\pm$0.01 &1.05\scriptsize$\pm$0.01\\\hline
    \end{tabular}
 \end{table}

Additional numerical experiments have been conducted to study other
choices of $\Kp$ and $\Kn$ for a Forward scheme and can be found in
the supplement, \Cref{sec:numer-comp-betw} and
\Cref{sec:numer-comp-diff}. They show that Forward Ref is one of the
most efficient tested schemes, that a direct Forward EC with a
full-orthogonal refreshment every $\rmT$ is similar to Forward Full
Ref and that sparse-orthogonal refreshment schemes are more robust to
the choice of $\rmT$ for the refreshment than a full-orthogonal one.

\subsection{Poisson-Gaussian Markov random field}

\label{sec:poisson-gauss}

To assess the performance for more complex models, we now
  consider a Poisson-Gaussian Markov random field model similarly to
  \cite[Section 4.5]{bouchard:vollmer:doucet:2016}. In this setting,
  the observations
  $\mathbf{Y} = (Y_{i,j})_{(i,j) \in \{1,\ldots,\tilde{d}\}^2} \in
  \nset^{\tilde{d}^2} $
  are supposed to be independent samples such that for any
  $(i,j) \in \{1,\ldots,\tilde{d}\}^2$, $Y_{i,j}$ has a Poisson
  distribution with parameter $\exp(x_{i,j})$ with $x \in \rset$. The
  parameter $\bfx = (x_{i,j})_{(i,j) \in \{1,\ldots,\tilde{d}\}^2}$ is
  assumed to be a Gaussian random field, \ie~the prior distribution is
  set to be the zero-mean Gaussian distribution with covariance matrix
  $\tilde{\Sigma}_{(i,j),(\tilde{i},\tilde{j})} =
  \exp(-(2\upsigma^2)^{-1}\{(i-j)^2+(\tilde{i}-\tilde{j})^2\}^{1/2})$,
  and $\upsigma^2 =\tilde{d}-1$.  The target distribution $\pi$ admits a potential $U$ given for any
  $\bfx \in \rset^{\tilde{d}^2}$ by
\begin{equation*}
  U(\bfx)  = \frac 12\sum_{i,j,\tilde{i},\tilde{j}=1}^{\tilde{d}} x_{i,j} \tilde{\Sigma}_{(i,j),(\tilde{i},\tilde{j})}^{-1}x_{\tilde{i},\tilde{j}} + \sum_{i,j=1}^{\tilde{d}}\{\exp(x_{i,j}) - Y_{i,j}x_{i,j}\}  \eqsp.
\end{equation*}
 
\begin{figure}[h]
\begin{center}
  \includegraphics[width=1.0\textwidth]{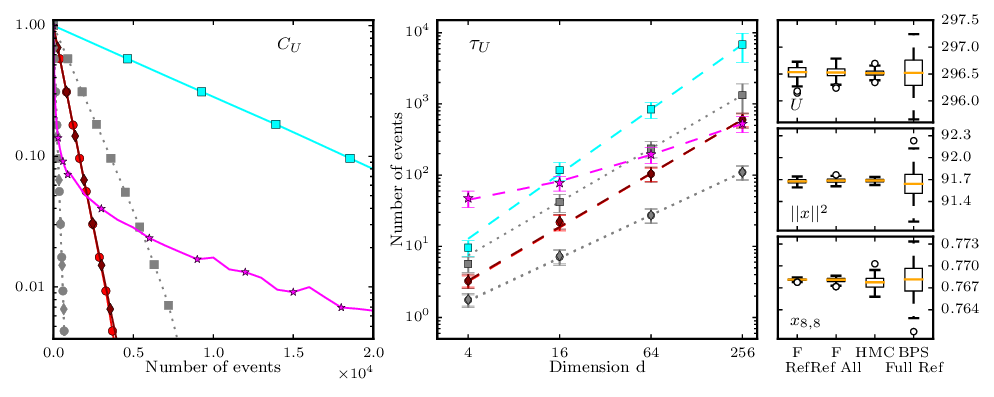}
\caption{ \footnotesize\textbf{Left:} Autocorrelation functions $C_U$
  for the potential U for $d=256$ for HMC (magenta, star), Forward Ref
  (red, circle), Forward Ref All (maroon, diamond) and BPS Full Ref
  (cyan, square). \textbf{Middle:} Scaling with $d$ of the integrated
  autocorrelation times $\tau$ of $U$ for the same schemes. For both
  $C_U$ and $\tau$, the time unit is the number of gradient
  evaluations or events. The jump times/true events of the PDMC are
  computed through thinning and required several fake events, results
  in terms of true events/jump times are represented with a dotted
  grey line and with the respective markers. \textbf{Right:} Boxplots
  for $U$, $\norm{x}^2$ and $x_{88}$ at $d=256$. For the box plots,
  samples are separated by the same averaged CPU times
  ($\sim 1.10^{-3}s$), resulting in average to 500 gradient
  evaluations for HMC and from 137 (Forward) to 155 (BPS) events for
  PDMC schemes per sample.}
\label{fig:markov_poisson_Num}
\end{center}
\end{figure}

We compared the schemes Forward Ref, Forward Ref All, HMC and BPS
  for the dimensions $d = \tilde{d}^2 \in \{4,16,64,256\}$. The
  refreshment time is set to $T =5$ for Forward Ref and to $T = 2$ for
  BPS. This choice achieves the fastest decorrelation in $U$, which
  appears to be the slowest observable to converge, in comparison to
  $\norm{\bfx}^2$ and $\bfx$. The event times of the underlying PDMP are
  computed through the decomposition and thinning of the target
  Poisson process. We refer to
  \Cref{sec:sampling-poisson} in the supplement for details on this
  procedure and to \cite{michel:kapfer:krauth:2014} and
  \cite{bouchard:vollmer:doucet:2016}.

\begin{table}
  \caption{\small  Scaling $z$ of the integrated autocorrelation times
    $\tau_U$ with the dimension $d$ in the Poisson-Gaussian Markov random field model, for the least-squares fit $\tau_U=Ad^z+B$ by Levenberg-Marquardt gradient method. For PDMC schemes, results are given in terms of  gradient evaluations/all events, jump times/true events and CPU times, for HMC, in terms of gradient evaluations/leapfrogs and CPU times.
    \label{tab:Scaling_poisson}}
\setlength\tabcolsep{0.1cm}
\small
\begin{center}
    \begin{tabular}{ccc|ccc|ccc|cc}
\hline
 \multicolumn{3}{c}{F. Ref}           &\multicolumn{3}{c}{F. Ref All}       &\multicolumn{3}{c}{BPS Full}  &     \multicolumn{2}{c}{HMC}     \\\hline
 \scriptsize All Ev.& \scriptsize True Ev. & \scriptsize CPU  & \scriptsize All Ev.& \scriptsize True Ev.&\scriptsize CPU&\scriptsize  All Ev.& \scriptsize True Ev.& \scriptsize CPU&\scriptsize Grad. Eval& \scriptsize CPU \\\hline
    \scriptsize    1.27\scriptsize$\pm$0.10 &\scriptsize 1.00 \scriptsize$\pm$0.05 &\scriptsize 2.4 \scriptsize$\pm$0.1 &\scriptsize 1.27\scriptsize$\pm$0.10 &\scriptsize 1.00 \scriptsize$\pm$0.05&\scriptsize 2.3 \scriptsize$\pm$0.1 &\scriptsize 1.5\scriptsize$\pm$0.1&\scriptsize 1.25 \scriptsize$\pm$0.10 &\scriptsize 3.1 \scriptsize$\pm$0.1 &\scriptsize 0.8\scriptsize$\pm$0.1&\scriptsize 2.3 \scriptsize$\pm$0.1\\\hline
    \end{tabular}
\end{center}
 \end{table}

\Cref{fig:markov_poisson_Num} displays the autocorrelation
  functions and the integrated autocorrelation times for the potential
  $U$ obtained from the different schemes. The fitted scaling of the
  latter with the dimension $d$ can be found in
  \Cref{tab:Scaling_poisson}. In addition,
  \Cref{fig:markov_poisson_Num} also exhibits boxplots for $U$, the
  norm $\norm{\bfx}^2$ and the component $x_{8,8}$ for $d = 256$.
  First, as the event times are computed through a thinning procedure,
  it leads to an extra computational cost compared to a direct
  computation, if it was available. Then, the results show that BPS is
  outperformed in all situation. Forward Ref and Forward Ref All
  behaves in the same manner, confirming that the refreshment tuning
  for Forward Ref schemes is also not crucial in that case. HMC, if
  displaying slower decorrelations than Forward-type schemes, shows a
  better scaling, except when given in CPU time. Finally, from the
  boxplots, Forward-type schemes seems to be the more efficient for
  the component $x_{8,8}$ and as efficient as HMC for the norm.

\subsection{Mixture of Gaussian distributions}
\label{sec:mixgaussian}

Our next numerical experiment is based on the sampling of a mixture of
$5$-Gaussian distributions of dimension $d$ to test whether a
direct-sampling scheme could lead to difficulties to get out of a
local mode. In order to introduce some randomness, a set of $d$ random
numbers $(\sigma^2_i)_{i \in \{1,\ldots,d\}}$ is picked uniformly
between $0.5$ and $3$. Then for $j \in \{1,\ldots, 5\}$, we consider
the Gaussian distribution with mean $\mu_j$ and covariance matrix
$\Sigma_j$ where
$\Sigma_j =
\diag(\sigma_{\kappa_j(1)}^2,\ldots,\sigma_{\kappa_j(d)}^2)$,
where $(\kappa_j)_{j \in \{1,\ldots,5\}}$ is a sequence of $5$
uniformly-random permutations of $\{1,\ldots,d\}$, therefore
$(\Sigma_j)_{j \in \{1,\ldots,N\}}$ are equal up to a rotation. The
mean are defined recursively by $\mu_j=0$ if $j=1$ and for $j >1$ by
$ \mu_j = \mu_{j-1} + (\nu_1\sigma_{\kappa_j(i)} +
\nu_2\sigma_{\kappa_{j-1}(i)})^{\top}_{i \in \{1,\ldots,d\}}$,
where $\nu_1, \nu_2$ are uniform samples between $1$ and $2$. This
choice has been made to ensure a separation between each mode of at
least both standard deviations. Each Gaussian distribution has equal
probability in the mixture. The event time can be computed through a
thinning procedure, as done in \cite[Example 2]{wu:robert:2017}.

\begin{figure}
\begin{center}
\includegraphics[width=1.0\textwidth]{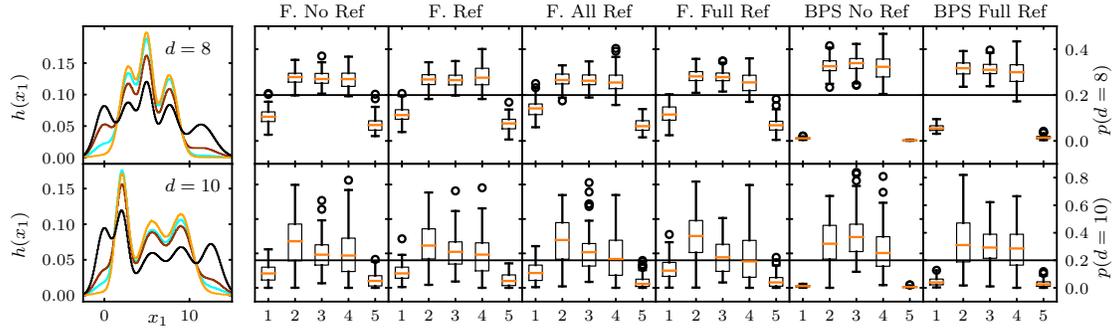}
\caption{ \footnotesize Histograms of component $x_1$ for the Gaussian
  Mixtures in dimension $d=8$ (\textbf{Top Left}) and $d=10$
  (\textbf{Bottom Right}) with, ordered from darker to lighter colors,
  real distributions (black), Forward Ref All (maroon), BPS No Ref
  (orange) and BPS Full Ref (cyan). The \textbf{right} panel shows the
  boxplots of the estimated mixture probabilities for $d=8$
  (\textbf{Top}) and $d=10$ (\textbf{Bottom}) for different schemes.}
\label{fig:Ran_Num_Hist}
\end{center}
\end{figure}

In small dimensions ($d=2,4,5$), all the tested schemes give similar
results in terms of efficiency, as exhibited by the ESS of the vector
$x$ and the square norm $\norm{x}^2$ in \Cref{tab:Ran_ESS} in the
supplementary. Contrary to the former Gaussian distribution, Forward
No Ref and BPS No Ref appear to be ergodic.  A sharp drop in the ESS
can be observed between $d=2$ and $d=4$, especially for BPS No
Ref, and, in higher dimensions ($d=8,10$), convergence is very
slow. Refreshment time is tuned to $\rmT=100$ ($\sim 40$ events). We
display in \Cref{fig:Ran_Num_Hist} the histograms of the first
component of $x$ averaged on $100$ runs of $4\times 10^6$ events and
started from random initial positions drawn from the real
distribution. Forward Ref All shows a better exploration than BPS Full
Ref and BPS No Ref. This result is confirmed by the boxplots of the
estimated mixture probabilities, obtained by assigning each successive
sample of each run to a distribution based on the closest mean. Here
runs were all started from $0$. A clear difference can be observed
between Forward schemes and BPS schemes, as they share inside their
class similar results in terms of exploration of the extreme modes,
despite different refreshment schemes. BPS Full Ref shows a better
exploration than BPS No Ref though. All in all, Forward methods do outperform
BPS ones, the most efficient being Forward Ref All.

\subsection{Logistic regression}
\label{sec:logreg}

We focus in this Section on a Bayesian logistic regression
problem. The data $(y_i)_{i \in \{1,\ldots,N\}}$, $N \in \nset^*$, are
assumed to be \iid~Bernouilli random variables with probability of
success $\mathrm{logit}(\ps{x_i}{ \theta})$ for any
$i \in \{1,\ldots,N\}$, where $(x_i)_{i \in \{1,\ldots,N\}}$ are
covariate variables, $\theta$ is the parameter of interest and
$\mathrm{logit}(u) = \rme^{u}/(1+\rme^{u})$, for any $u \in
\rset$.
The prior distribution on $\theta$ is assumed to be the
$d$-dimensional zero-mean Gaussian distribution with covariance matrix
$1000 \times \Idd$. Then, the a posteriori distribution for this model
has a potential given for any $\theta \in \rset^d$ by
  \begin{equation}
    \label{eq:U_log_reg}
    U(\theta)  = \sum_{i=1}^N \defEns{-y_i\ps{x_i}{\theta}  + \log\parentheseDeux{1+\exp(\ps{x_i}{\theta})}} + \norm[2]{\theta}/(2 \varsigma^2 ) \eqsp, \, \varsigma^2 = 1000 \eqsp. 
  \end{equation}
  We perform our numerical studies on the German credit dataset
  ($N=1000$, $d=25$) and Musk dataset ($N=476$, $d=167$) from the UCI
  repository \cite{Dua:2017}.  The procedure we follow for this
  example has been proposed in \cite{michel:kapfer:krauth:2014} and
  \cite[Section 3]{bouchard:vollmer:doucet:2016}, and uses a
  decomposition of the gradient of the potential over the data. We
  refer to \Cref{sec:DPDMC} in the supplement for more details.

\begin{figure}[h!]
\begin{center}
\includegraphics[width=1.0\textwidth]{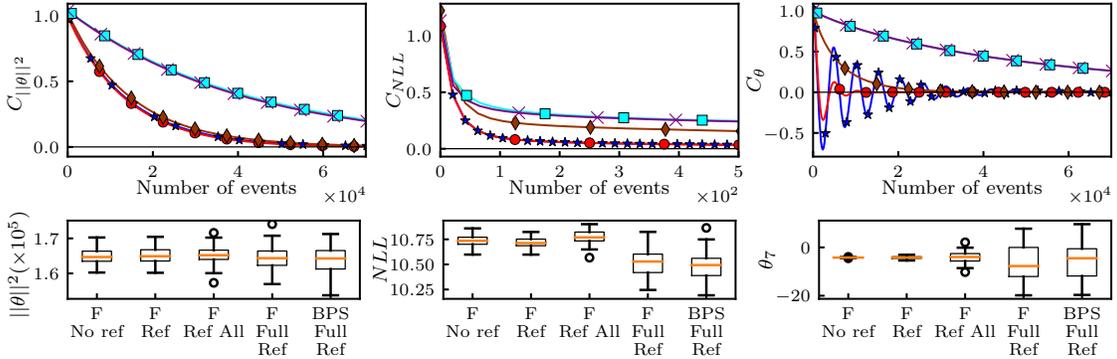}
\caption{ \footnotesize\textbf{Top:} Autocorrelation functions for the
  squared norm of the weights $\norm{\theta}^2$ (\textbf{left}), the
  negative loglikelihood (NLL, \textbf{middle}) and $\theta$
  (\textbf{right}) for the Musk dataset for Forward No Ref (blue,
  star), Forward Ref (red, circle), Forward Ref All (maroon, diamond),
  Forward Full Ref (purple, cross) and BPS Full Ref (cyan,
  square). \textbf{Bottom:} Boxplots for $\norm{\theta}^2$, NLL and
  $\theta_{7}$, component with the highest variance.  }
\label{fig:LogRegM_Num}
\end{center}
\end{figure}

\begin{figure}[h!]
\begin{center}
\includegraphics[width=1.0\textwidth]{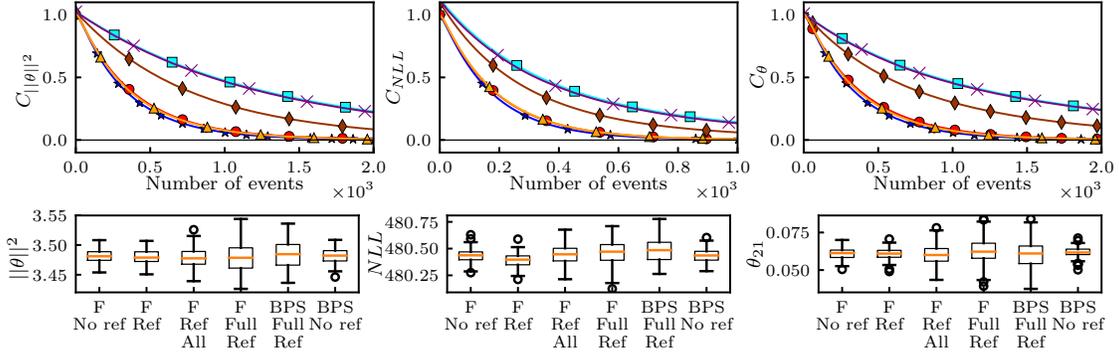}
\caption{ \footnotesize\textbf{Top:} Autocorrelation functions for the
  squared norm of the weights $\norm{\theta}^2$ (\textbf{left}), the
  negative loglikelihood (NLL, \textbf{middle}) and $\theta$
  (\textbf{right}) for German Credit dataset for Forward No Ref (blue,
  star), Forward Ref (red, circle), Forward Ref All (maroon, diamond),
  Forward Full Ref (purple, cross), BPS Full Ref (cyan, square) and
  BPS No Ref (yellow, up triangle). \textbf{Bottom:} Boxplots for
  $\norm{\theta}^2$, NLL and $\theta_{21}$, component with the highest
  variance.}
\label{fig:LogRegG_Num}
\end{center}
\end{figure}

\begin{table}
\caption{ \footnotesize ESS for the Musk and German credit datasets per event.}
\small
{\begin{tabular}{l|ccc|ccc}
 \multicolumn{1}{c}{}&\multicolumn{3}{c}{Musk dataset}&\multicolumn{3}{c}{German credit dataset}\\[2pt]
Algorithm        & $\theta$     & $\norm{\theta}^2$ & NLL      &  $\theta$    & $\norm{\theta}^2$ & NLL \\\hline 
Forward No Ref   & 240\scriptsize$\pm$80   & 4.3\scriptsize$\pm$0.1    & 228\scriptsize$\pm$10    &160.0\scriptsize$\pm$0.8 &145\scriptsize$\pm$3    &324\scriptsize$\pm$5  \\
Forward Ref      & 128\scriptsize$\pm$6    & 4.57\scriptsize$\pm$0.09  & 234\scriptsize$\pm$10    &140.2\scriptsize$\pm$0.8 &137\scriptsize$\pm$3    &288\scriptsize$\pm$5   \\
Forward Ref All  & 9.0\scriptsize$\pm$0.1  & 3.7\scriptsize$\pm$0.1    & 77\scriptsize$\pm$4      &64.9\scriptsize$\pm$0.7  & 73\scriptsize$\pm$2    &157\scriptsize$\pm$4   \\
Forward Full Ref & 1.03\scriptsize$\pm$0.02& 1.23\scriptsize$\pm$0.08  & 20\scriptsize$\pm$1      &34.5\scriptsize$\pm$0.9  &44\scriptsize$\pm$1     &107\scriptsize$\pm$3   \\
BPS No Ref       & --           & --             & --            &149.7\scriptsize$\pm$0.7 &143\scriptsize$\pm$3    &282\scriptsize$\pm$5   \\
BPS Full Ref     & 1.07\scriptsize$\pm$0.02& 1.23\scriptsize$\pm$0.07  &  20\scriptsize$\pm$1     &35.4\scriptsize$\pm$0.7  &40\scriptsize$\pm$2     &96\scriptsize$\pm$3\\\hline
\end{tabular}}\\
\footnotesize NOTE: All results are multiplied by $10^5$.
\label{tab:LogReg_ESS}
\end{table}

The refreshment time is fixed to $\rmT=10$ for the Musk dataset,
corresponding to an average of $22$ events for Full Ref schemes, $21$
otherwise and to $\rmT=0.1$ for the German Credit
dataset,
corresponding to an average of $13$ events for Full Ref schemes, $12$
otherwise. Both BPS No Ref and Forward No Ref were also tested
  but BPS No Ref appears not to be ergodic on the Musk dataset,
  whereas no ergodicity issue were encountered with Forward No Ref.
The autocorrelation functions for $\norm{\theta}^2$, the negative
loglikelihood and $\theta$ are shown in Figure~\ref{fig:LogRegM_Num}
for the Musk dataset and in Figure~\ref{fig:LogRegG_Num} for the
German Credit dataset. Forward Full Ref and BPS Full Ref have matching
decorrelations on both datasets, as Forward Ref and Forward No Ref
(except for the decorrelation of $\theta$ for the Musk dataset). We
can observe that Forward schemes based on an orthogonal switch or no
refreshment are faster and display their robustness to the
refreshment-time tuning, as the decorrelation decay is always stronger
from Forward No Ref to Forward Ref All. Quantitative accelerations can
be found by comparing the ESS summarized in \Cref{tab:LogReg_ESS}.

\section{Conclusion}
\label{sec:conclu}

In this paper, we have introduced a generalized class of PDMC, Forward
event-chain Monte Carlo, by exploiting the rotation symmetry around
the gradient and relying on a global stochastic picture. The main
practical asset is its flexibility, as it gives new possibilities in
terms of refreshment and new-direction sampling schemes, while
  no requiring any crucial fine-tuning. By breaking free from quasi
iso-potential trajectories, it allows to reduce the need for extra
randomization and improves the efficiency of the exploration. Our
numerical experiments also show that standard PDMC
benefit from only transitioning to the Forward global
stochastic direction sampling, while keeping the same refreshment
scheme. Again, we stress that the stochastic direction
  sampling at event is required for correctness and is not equivalent
  to some artificial refreshment, the latter easily leading to a
  random-walk behavior.

  In practice, we presented a collection of refreshment and
  new-direction sampling schemes which proved to bring accelerations
  in practice. There are however many possible other choices and a
  promising research axis lies in a quantitative theoretical study of
  the optimal refreshment scenario, depending on the problem at
  hand. A first question is how one can use the target geometry to
  determine which kernels to implement locally. Another line of
    research is to explore the impact of different choices of
    differential flows $\phi$ and function rates $\lambda$ than the
    ones considered here. To conclude, while PDMC appear as an
  exciting new MCMC development, a complete theoretical understanding
  is lacking and a key issue is indeed how to find the correct trade
  off between a diffusive exploration and a non-ergodic ballistic one.

\section*{Acknowledgements}
M. M. is very grateful for the support from the Data Science
Initiative and M. M and A. D. thanks the Chaire BayeScale
"P. Laffitte" for its support. We are grateful to the cluster
computing facilities at \'Ecole Polytechnique (mésocentre PHYMATH) and
the Mésocentre Clermont Auvergne University for providing computing
resources.

\bigskip

\begin{center}
{\large\bf SUPPLEMENTARY MATERIAL}
\end{center}

\begin{description}
\item[Appendix A Formal derivation of new Markov kernels $Q$]
  Complementary details to the presentation in Section
  \ref{sec:derivation-new-pdmps}. (pdf)
\item[Appendix B Choices of $\Kp$]
  Expression for Metropolis-Hastings choice for $\Kp$. (pdf)
\item[Appendices C1 to C4 Implementation details] Presentation and pseudocodes
  of the extended PDMP MCMC process relying on a fixed-time
  refreshement. Additional technical details on event-time
  computations. Presentation and pseudocode of distribution
  factorization for PDMC implementation. (pdf)
\item[Appendices D1 to D5 Addition to the numerical experiments]
  Further numerical results on comparaison to a Hastings-Metropolis
  scheme for the anisotropic Gaussian distribution, choices of $\Kp$, $\Kn$,
  $\rmT$, $\rmT$, $p$ and $\theta$, ESS for mixture of Gaussian
  distributions. (pdf)

\end{description}

\newpage
\appendix

\section{Formal derivation of new Markov kernels $Q$}
\label{sec:form-deriv-new}

We introduce the extended generator $\generator$ (see \cite[Theorem
26.14]{davis:1993}) associated with the PDMP semi-group
$(P_t)_{t \geq 0}$ described in \Cref{sec:forward-p-orthogonal} with
$\varphi$ and $\rate$ given by \eqref{eq:differential_flow} and
\eqref{eq:definition_rate}.  It is
the operator $\generator$ defined for all $f \in \mrc^1_c(\rset^{2d})$
and $(x,y) \in \rset^{2d}$ by
\begin{multline}
  \label{eq:definition_generator_rough}
  \generator f(x,y) = \ps{y}{\nabla_x f(x,y)} + \ps{y}{\nabla U(x)}_+ \defEns{\int_{\rset^d}f(x,\tilde{y}) Q((x,y),\rmd \ty) - f(x,y) } \\
  + \lambdab\defEns{\int_{\rset^d} f(x,\ty) \rmd \loiy(\ty)- f(x,y)}
  \eqsp,
\end{multline}
where $\mrc^1_c(\rset^{2d})$ are the set of differentiable function
from $\rset^{2d}$ to $\rset$ with compact support.  Then by
\cite[Proposition 9.2]{ethier:kurtz:1986}, for $\tilde{\pi}$ to be an
invariant measure for $(X_t,Y_t)_{t \geq 0}$, it turns out that it is
necessary that for all $f \in \mrc^1_c(\rset^{2d})$,
\begin{equation*}
  \int_{\rset^{d} \times \rset^d} \generator f(x,y) \rme^{-U(x)} \rmd x \, \rmd \loiy(y) = 0 \eqsp.
\end{equation*}
By integration by part and elementary algebra based on
\eqref{eq:definition_rate}, \eqref{eq:definition_kernelKQ} and
\eqref{eq:definition_generator_rough}, this condition is equivalent to,
for all $f  \in \mrc^1_c(\rset^{2d})$
\begin{multline*}
  \int_{\rset^d \times \rset^d} \int_{\rset^d} \ps{y}{\nabla U(x)}_+
  f(x,\ty) \kernelQ((x,y),\rmd \ty) \rme^{-U(x)} \rmd x \rmd \loiy(y)
  \\= \int_{\rset^d \times \rset^d}\ps{\ty}{\nabla U(x)}_- f(x,\ty)
  \rme^{-U(x)} \rmd x \rmd \loiy(\ty) \eqsp.
\end{multline*}
Furthermore, using Fubini's theorem, this relation holds if the following
condition is satisfied for almost all $x \in \rset^d$,
\begin{equation}
\label{eq:extended_global_balance}
  \int_{\rset^d} \int_{\rset^d} \ps{y}{\nabla U(x)}_+ f(x,\ty) \kernelQ((x,y),\rmd \ty)  \rmd \loiy(y) =  \int_{\rset^d} \ps{\ty}{\nabla U(x)}_- f(x,\ty)  \rmd \loiy(\ty) \eqsp.
\end{equation}

As $\loiy$ is rotation invariant, \eqref{eq:extended_global_balance}
is equivalent to the fact that for all $x \in \rset^d$, such that
$\nabla U(x) \not = 0$, the probability measure defined for all
$\msa \in \mcb{\rset^d}$ by
\begin{equation}
  \label{eq:def_mupp}
\mupp^{x}(\msa) =  \left. \int_{\rset^d} \1_\msa(\tilde{y}) \ps{\ty}{\nabla U(x)}_- \loiy(\rmd \ty) \middle/  \int_{\rset^d}  \ps{\ty}{\nabla U(x)}_- \loiy(\rmd \ty) \right. \eqsp,
\end{equation}
is invariant for the Markov kernel on $\rset^d \times \mcb{\rset^d}$, $(y,\msa) \mapsto R Q ((x,y),\msa)$ where $R$ is the extended Reflection kernel on $\rset^{2d} \times \mcb{\rset^{2d}}$ defined for all $(x,y) \in \rset^{2d}$ by
$ R((x,y),\cdot) = \updelta_{(x,\phi_R(x,y))}(\cdot)$, where $\phi_R$  is defined in \eqref{eq:bouncy_kernel}.
Simply put, $\mupp^x$ is the \emph{reflected-event} distribution, the
probability distribution for the reflection of the new direction to
trigger an event. From this observation, we derive a necessary general
expression for the Markov kernel $\Qkernel$. In
\eqref{eq:extended_global_balance}, $x \in \rset^d$ is assumed to be
fixed and if $\nabla U(x) = 0$, then any choice of $\Qkernel$ is
suitable and we choose $\Qkernel((x,y),\cdot) = \loiy$. We consider
now the case $x \in \rset^d$ satisfies $\nabla U(x) \not = 0$. The
projection on $\nabla U(x)$ is essential in
\eqref{eq:extended_global_balance} and that is why we disintegrate
$\loiy$ accordingly. A global symmetry around $\nabla U$ then appears
and circumvents any introduction of additional symmetry. More
precisely, we define the map $\projp : \rset^d \to \rset$,
$\projn : \rset^d \to \spana(\nabla U(x))^{\perp}$ for all
$y \in \rset^d$ by $\projp(y) = \ps{y}{\mrn(x)}$, and
$\projn(y) = \{\Id-\rmna{x} \rmn(x) ^{\transpose}\}y$.  In addition, consider the pushforward measure of $\loiy$
by $\projp$ given for all $\msa \in \mcb{\rset}$ by
\begin{equation}
  \label{def:loiyp}
  \loiyp(\msa) = \loiy\defEns{\left(\projp\right)^{-1}\parenthese{\msa}} \eqsp.
\end{equation}
Since $\loiy$ is rotation invariant, $\loiyp$ does not depend on $x$.
Let $T^{x,\perp}$ be the regular conditional
distribution of $\loiy$ given $\projp$
\cite[Theorem 10.5.6.]{bogachev:2006}, defined on
$\rset \times \mcb{\rset^{d}}$ such that for all $\msa \in \rset^d$,
\begin{equation}
\label{eq:decomposition_muy}
  \loiy(\msa) = \int_{\rset^{d+1}}\1_{\msa}\parenthese{y_{\sslash} \rmna{x}+ \projn(y'_{\perp})} T^{x,\perp}(y_{\sslash},\rmd y'_{\perp}) \mrd \loiyp(y_{\sslash}) \eqsp.
\end{equation}
Then, we assume that $\Qkernel$ in \eqref{eq:definition_kernelKQ} can be decomposed as follows for all
$(x,y) \in \rset^{2d}$ and $\msa \in \mcb{\rset^d}$,
\begin{multline}
  \label{eq:decomposition_Q_1}
  \Qkernel((x,y) , \msa) =  \int_{\rset^{d+1}} \1_{\msa}\parenthese{\ty_{\sslash} \rmna{x}+ \projn(\ty'_{\perp})} \\
  \Qp((x,\projp(y)),\rmd \ty_{\sslash})  \Qn\parenthese{(x,\ty_{\sslash},\projn(y)),\rmd \ty'_{\perp} }\eqsp,
\end{multline}
where $\Qn$ and $\Qp$ are Markov kernels on
$\rset^{2d} \times \mcb{\rset^{d}}$ and
$\rset^{d+1} \times \mcb{\rset}$ respectively.  

As illustrated on \Cref{fig:NewPick}, the decomposition divides the
Markov kernel $\Qkernel$ into two Markov kernels $\Qp$ and $\Qn$
which, starting from $(x,y)$, give a new direction $\ty$ by choosing
the new component of $\ty$ along $\nabla U(x)$ and the components of
$\ty$ in $\spana(\nabla U(x))^{\perp}$ respectively. In other word,
the sampling of $\Qkernel$ can then be decomposed in three steps:
starting from $(x,y) \in \rset^{2d}$
  \begin{enumerate}
  \item  sample $\tilde{Y}_{\sslash}$ from the probability measure on
  $\rset$, $\Qp((x,\projp(y)),\cdot)$;
\item sample $Y'_{\perp}$ from $\Qn\parentheseLigne{(x,\tilde{Y}_{\sslash},\projn(y)),\cdot}$ and set $\tilde{Y}_{\perp}=\projn(Y'_{\perp})$;
\item set $\tilde{Y}_{\sslash} \mrn(\nabla U(x)) + \tilde{Y}_{\perp}$, as the new direction.
  \end{enumerate}

  In the following, we establish sufficient conditions on $\Qp$ and
  $\Qn$ which imply that $\mupp^x$ given by \eqref{eq:def_mupp} is
  invariant with respect to
  $(y,\msa) \mapsto R Q ((x,y),\msa)$, which, as noticed,
  implies in turn that $\tpi$ defined by \eqref{eq:extended_target}
  is invariant for the PDMP Markov semi-group $(P_t)_{t \geq 0}$
  defined by $(\varphi,\rate,M)$ given in
  \eqref{eq:differential_flow}-\eqref{eq:definition_rate} and
  \eqref{eq:definition_kernelKQ}.
  
  Based on \eqref{eq:decomposition_Q_1}, since $\mupp^x$ given by \eqref{eq:def_mupp} has to be invariant
with respect to $(y,\msa) \mapsto R Q ((x,y),\msa)$, we have
that necessarily the pushforward measure of $\mupp^x$ by $\projp$ is
invariant with respect to the Markov kernel $\Kp$ on $\rset \times \mcb{\rset}$ given for $\yp \in \rset$ and $\msa \in \mcb{\rset}$ by
\begin{equation}
  \label{eq:def_tK_p}
\Kp(\yp,\msa) =  \Qp((x,-\yp),\msa)   \eqsp,
\end{equation}
 or
equivalently since $\loiy$ is rotation invariant that
$\Kp$ has $\loiym$ for invariant probability
measure on $(\rset,\mcb{\rset})$,  defined for all $\msa \in \rset$ by 
\begin{equation}
  \label{eq:def_loiy_sslash}
  \loiym(\msa) =   \left. \int_{\rset} \1_{\msa}(\yp) \parenthese{\yp}_-\loiyp(\rmd \yp) \middle/ \int_{\rset}  \parenthese{\yp}_-\loiyp(\rmd \yp) \right.  \eqsp,
\end{equation}
where $\loiyp$ is defined in \eqref{def:loiyp}.
As the Markov kernel $S(\yp,\msa) = \updelta_{-\yp}(\msa)$ is an involution, $S^2 = \Id$, it defines a one-to-one correspondence between Markov kernels $K$ leaving $  \loiym$ invariant and Markov kernels $\tilde{K}$ such that $(\yp,\msa) \mapsto \tilde{K}(-\yp,\msa)$ leaves $\loiym$ invariant. Then, the choice of $\Kp$ determines $\Qp$.
This observation sets the contribution of $Q$ along the direction $\nabla U(x)$.

By \eqref{eq:def_mupp} and \eqref{eq:decomposition_muy}, another condition  to ensure that $\mupp^x = \mupp^x R Q$ if the decomposition \eqref{eq:decomposition_Q_1} holds,  is that for any $\msa \in \mcb{\rset^d}$, $x \in \rset^d$ and $y_{\sslash} \in \msy_1$,
\begin{multline}
  \label{eq:condition_tilde_q}
  \int_{\rset^d \times \rset^d} \1_{\msa}\parenthese{\yp \rmna{x} + \projn(\ty)} T^{x,\perp}(\yp,\rmd \yprime) \Qn((x,\yp,\projn(\yprime)),\rmd \ty) \\
  =  \int_{\rset^d} \1_{\msa}\parenthese{\yp \rmna{x} + \projn(\yprime)} T^{x,\perp}(\yp,\rmd \yprime) \eqsp,
\end{multline}
where $\msy_1$ is the support of $\loiyp$. 
This relation is equivalent to the fact that for any $x \in \rset^d$ and $y_{\sslash} \in \msy_1$ the pushforward measure of $T^{x,\perp}(\yp,\cdot)$ by $\projn$,
\begin{equation}
  \label{eq:def_tilde_T}
\tilde{T}^{x,\perp}(\yp,\msa) = T^{x,\perp}(\yp,(\projn)^{-1}(\msa)) \eqsp,  
\end{equation}
 defined on $\mcbb(\spana(\nabla U(x))^{\perp})$, is invariant for the Markov kernel $\Kn$ on $\spana(\nabla U(x))^{\perp} \times \mcbb(\spana(\nabla U(x))^{\perp})$ defined by
 \begin{equation}
   \label{eq:def_tilde_Kn}
  \Kn(\yn,\msa) = \int_{\rset^d} \1_{\msa}\parenthese{\projn(\ty)} \Qn((x,\yp,\yn),\rmd \ty) \eqsp,
\end{equation}
which determines the contribution of $\Qn$ in $\Qkernel$.
 
Finally, we end up with the following result.
\begin{theorem}
  \label{theo:invariance_BPS_generalized}
  Consider the PDMP semi-group $(P_t)_{t \geq 0}$ associated with the flow $\varphi$, the event rate $\lambda$ and the Markov kernel defined by \eqref{eq:differential_flow}, \eqref{eq:definition_rate} and \eqref{eq:definition_kernelKQ} and assume that $Q$ is on the form \eqref{eq:decomposition_Q_1} such 
  that for any $x \in \rset^d$,
  \begin{enumerate}[label=(\roman*)]
  \item   $\loiym$, given by \eqref{eq:def_loiy_sslash} is invariant for $\Kp$ defined by \eqref{eq:def_tK_p};
  \item for any  $\yp \in \rset$, $\tilde{T}^{x,\perp}(\yp,\cdot)$ given by \eqref{eq:def_tilde_T} is invariant for $\Kn$ given by \eqref{eq:def_tilde_Kn}.
  \end{enumerate}
 Then $\tpi$ is invariant for $(P_t)_{t \geq 0}$. 
\end{theorem}

\section{Choices of  $\Kp$}
\label{sec:choices-kp}
By
\eqref{eq:def_loiy_para_sphere}, the Markov kernels associated to the random walk or independent Metropolis-Hastings  algorithm on $\ccint{-1,0}$ with Gaussian or
uniform noise are 
defined respectively for any $\yp \in \ccint{-1,0}$ and $\msa \in \mcb{\rset}$ by
\begin{multline*}
  \Kp(\yp,\msa) \\
    = \int_{\rset} \1_{\msa}(-\fracp{-\yp+\sigma^2 v}) \min\parenthese{1, \frac{ \fracp{-\yp+\sigma^2 v} (1-(\fracp{-\yp+\sigma^2 v})^{2})^{(d-3)/2}}{ (-\yp) (1-\yp^{2})^{(d-3)/2}}} q(v) \rmd v \\
  + \updelta_{\yp}(\msa) \int_{\rset} \parentheseDeux{1- \min\parenthese{1, \frac{ \fracp{-\yp+\sigma^2 v} (1-(\fracp{-\yp+\sigma^2 v})^{2})^{(d-3)/2}}{ (-\yp) (1-\yp^{2})^{(d-3)/2}}}} q(v) \rmd v 
   \eqsp,
\end{multline*}
and
\begin{multline*}
  \Kp(\yp,\msa) \\
    = \int_{\rset} \1_{\msa \cap\ccint{-1,0}}(-\sigma^2 v) \min\parenthese{1, \frac{ q(\yp) \fracp{\sigma^2 v} (1-(\fracp{\sigma^2 v})^{2})^{(d-3)/2}}{ q(v) (-\yp) (1-\yp^{2})^{(d-3)/2}}} q(v) \rmd v \\
  + \updelta_{\yp}(\msa) \int_{\rset} \parentheseDeux{1- \min\parenthese{1, \frac{ q(\yp) \fracp{-\yp+\sigma^2 v} (1-(\fracp{-\yp+\sigma^2 v})^{2})^{(d-3)/2}}{ q(v) (-\yp) (1-\yp^{2})^{(d-3)/2}}}} q(v) \rmd v 
  \eqsp,
\end{multline*}
where $\sigma^2 >0$, $\fracp{t}$ is the fractional part of $t \in \rset$, and $q$ is for example  either the uniform distribution density on $\ccint{-1,1}$ or the standard Gaussian density. 
\section{Implementation Details}

\subsection{Details on refreshment strategy}
\label{sec:refresh_strat_app}

After extension of the state space $\mse = \rset^d\times \msv$ to
$\mse \times \{0,1\}$, we consider the following PDMP
$(\bX_t,\bY_t,\bB_t)_{t \geq 0}$ defined in
\Cref{algo:Ext_Forward_MCMC}, with $Q_0,Q_1$ two Markov kernels on
$\rset^{2d} \times \mcb{\rset^d}$ associated with
$\Kn_0,\Kn_1,\Kp_0,\Kp_1$ satisfying the conditions of
\Cref{theo:invariance_BPS_generalized}. This PDMP corresponds to the
differential flow $\bvarphi_t(x,y,b) = (x+ty,y,b)$ for any $(x,y,b)
\in \mse \times \{0,1\}$, event rate $\lambda$ given by
\eqref{eq:definition_rate} with $\lambdab=0$ and Markov kernel $\bQ$
on $(\mse \times \{0,1\} ) \times \mcb{\msy \times \{0,1\}}$ defined
for any $(x,y,b) \in \mse \times \{0,1\}$, $\msa \in \mcb{\msy}$ by
$\bQ((x,y,b),\msa\times \{1\}) = Q_b((x,y),\msa).$

\begin{algorithm}[h!]
 \caption{ \label{algo:Ext_Forward_MCMC} Extended PDMP MCMC process }
  \KwData{Markov kernels $Q_0,Q_1$  and initial points $(\bX_0,\bY_0,\bB_0) \in \rset^{2d} \times \{0,1\}$}
 \KwResult{Extended PDMC  $(\bX_t,\bY_t,\bB_t)_{t \geq 0}$ based on \Cref{theo:invariance_BPS_generalized}}
 Initialize $S_0 =0$ and a sequence of \iid~exponential random variables $(E_i)_{i \geq 1}$ with parameter $1$\\
 \For{$n \geq 0$}{
   \Indm   \mbox{Set $T_{n+1} = \inf\{ t \geq 0 \, : \, \int_{0}^t \ps{\bY_{S_n}}{\nabla U(\bX_{S_n} + u \bY_{S_n})}_+ \rmd u \geq E^1_{n+1}\}$}\\
      Set $S_{n+1} = S_n + T_{n+1}$\\
Set   $\bY_t = \bY_{S_n}$,  $\bX_{t} = \bX_{S_n} + (t-S_n) \bY_{S_n}$, $\bB_t = \bB_{S_n}$ for $t \in \ooint{S_n,S_{n+1}}$,
        $\bX_{S_{n+1}} = \bX_{S_n} + T_{n+1} \bY_{S_n}$\\
   \eIf{$\bB_{S_n}=0$}{
     Set $\bB_{S_{n+1}} = 1$ and sample $\bY_{S_{n+1}} \sim Q_0((\bY_{S_n},\bX_{S_{n+1}}),\cdot)$
   }
   {
     Set $\bB_{S_{n+1}} = 1$ and sample $\bY_{S_{n+1}} \sim Q_1((\bY_{S_n},\bX_{S_{n+1}}),\cdot)$
   }
   }
\end{algorithm}

Therefore, the generator associated with this PDMP is given for any
$f: \mse \times \{0,1\} \to \rset$ such that for any $i \in \{0,1\}$,
$f(\cdot,\cdot,i) \in \rmC^1_c(\rset^{2d})$, by
\begin{equation*}
  \bgenerator f(x,y,b) = \ps{y}{\nabla_x f(x,y,b)} + \ps{y}{\nabla U(x)}_+ \defEns{\int_{\rset^d}f(x,\tilde{y},1) Q_b((x,y),\rmd \ty) - f(x,y,b) } \eqsp,
\end{equation*}
for any $(x,y,b) \in \mse \times \{0,1\}$.  Using the same reasoning
as for the proof of \Cref{theo:invariance_BPS_generalized} and since
$\Kn_0,\Kn_1,\Kp_0,\Kp_1$ are assumed to satisfy the conditions of
this Theorem, it follows that for any $f: \mse \times \{0,1\} \to
\rset$ such that for any $i \in \{0,1\}$, $f(\cdot,\cdot,i) \in
\rmC^1_c(\rset^{2d})$, $\int_{\mse \times \{0,1\}} f(x,y,b) \rmd
\tilde{\pi}\otimes\updelta_{1} (x,y,b) = 0$ and that
$\tilde{\pi}\otimes\updelta_{1} $ is invariant for the semi-group
$(\bP_t)_{t \geq 0}$ associated with $(\bX_t,\bY_t,\bB_t)_{t \geq 0}$,
where $\tpi$ is given by \eqref{eq:extended_target}.  Therefore, we
can choose for $Q_0$ a Markov kernel associated with a kernel $\Kn_0$
different from the identity and $Q_1$ a Markov kernel associated with
a kernel $\Kn_1 = \Id$. However, the problem is that once $Q_0$ is
used, the extra variable is permanently set to $1$ and $Q_1$ is always
used in the sequel of the algorithm.  The idea is then to fix a time
$\rmT >0$ and update the extra variable $(\bB_t)_{t \geq 0}$ to $0$ to
use once again $Q_0$ and therefore partially refresh the
direction. This procedure is described in
\Cref{algo:update_Ext_Forward_MCMC}. While the full process
$(\bX_t,\bY_t,\bB_t)_{t \geq 0}$ cannot be ergodic, it seems
numerically that $(\bX_t,\bY_t)$ is ergodic with respect to $\tpi$
defined by \eqref{eq:extended_target}.

\begin{algorithm}[h!]
 \caption{ \label{algo:update_Ext_Forward_MCMC} Extended PDMP MCMC process with refreshment}
  \KwData{Markov kernels $Q_0,Q_1$, initial points $(\bX_0,\bY_0,\bB_0) \in \rset^{2d} \times \{0,1\}$ and a time $\rmT >0$}
 \KwResult{Refreshed extended PDMC  $(\bX_t,\bY_t,\bB_t)_{t \geq 0}$ based on \Cref{theo:invariance_BPS_generalized}}
 \For{$n \geq 0$}{
   For $t \in \ccint{n \rmT, (n+1)\rmT}$ sample $\bX_t,\bY_t,\bB_t$ according to \Cref{algo:Ext_Forward_MCMC}\\
   Set $\bB_{(n+1)\rmT} = 0$
   }
\end{algorithm}

\begin{figure}
\begin{center}
 \includegraphics{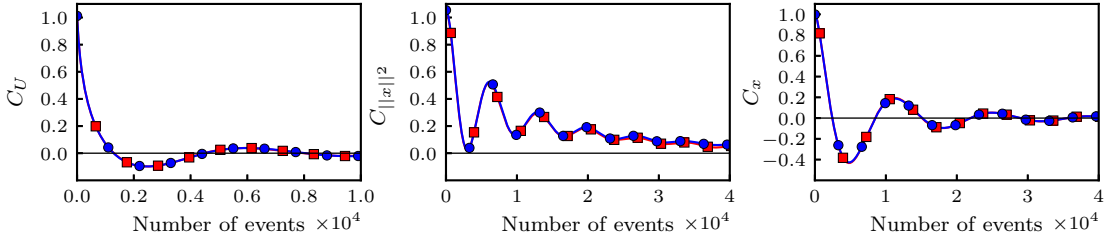}
\caption{Autocorrelation functions $C$ of the potential $U$
  (\textbf{Left}), the squared norm $||x||^2$ (\textbf{Middle}) and
  $x$ (\textbf{Right}) for the ill-conditioned zero-mean Gaussian
  distribution with covariance matrix given by \eqref{eq:Gaussian} and
  $d=400$ for direct Forward EC schemes with a fixed-time orthogonal
  switch (blue, circle) or to a stochastic orthogonal switch (red,
  square). \label{fig:Gauss_Num_FixMix}}
\end{center}
\end{figure}

We compare the described strategy with the mixture associated to the
identity kernel given in \Cref{sec:refresh_strat} on sampling the
zero-mean Gaussian distribution of
\Cref{sec:gaussian}. Fig. \ref{fig:Gauss_Num_FixMix} displays the
autocorrelation functions of $U$, $||x||^2$ and $x$ for $d=400$ for
either a choice of $\Kn$ set to a mixture parametrized by $p_r=0.018$ of
$\Id$ and of an orthogonal switch and one set to an orthogonal switch
every fixed time $\rmT=500$ and to $\Id$ otherwise. Both choices, with
these values of $p_r$ and $\rmT$, lead to the same averaged number of
events between two orthogonal refreshment ($\sim 55$). As can be seen
on Fig. \ref{fig:Gauss_Num_FixMix}, the same decorrelation is
obtained.

The stochastic refreshment step ruled by the Poisson process of rate
$\lambdab$ can be transformed into a refreshment process happening
also at every time $\rmT$. To do so, one needs to consider a set of
successive PDMPs with refreshment rate $\lambdab=0$, instead of a
single one, as described in \Cref{algo:chain_length}. When $\loiy$ is
the uniform distribution over the $d$-dimensional sphere $\sphere^{d-1}$,
$\rmT$ is also referred to as the \emph{chain length} in the physics
literature \citep{bernard:2009,michel:kapfer:krauth:2014}, where
fixing the refreshment time to $\rmT$ is particularly useful, e.g. for
particle systems with periodic boundary conditions.
\begin{algorithm}
 \caption{ \label{algo:chain_length} Implementation by a discrete collection of PDMP}
  \KwData{Markov kernels $Q$, initial points $(\bX_0,\bY_0) \in \rset^{2d}$ and a time $\rmT >0$}
 \KwResult{Generic PDMC through a collection of PDMP  $(\bX_t,\bY_t)_{t \geq 0}$ based on \Cref{theo:invariance_BPS_generalized}} 
Initialize $S_0 =0$, $T^{\text{Ref}}_0=T$ and a sequence of \iid~exponential random variables $(E_i)_{i \geq 1}$ with parameter $1$\\
 \For{$n \geq 0$}{
 \mbox{Set $T^{\text{Ev}}_{n+1} = \inf\{ t \geq 0 \, : \, \int_{0}^t \ps{Y_{S_n}}{\nabla U(X_{S_n} + u Y_{S_n})}_+ \rmd u \geq E_{n+1}\}$}\\
  \mbox{Set $T_{n+1} = \min(T_{n+1}^{\text{Ev}},T^{\text{Ref}}_{n+1})$ and   $S_{n+1} = S_n + T_{n+1}$}\\
  \mbox{Set   $Y_t = Y_{S_n}$,  $X_{t} = X_{S_n} + (t-S_n) Y_{S_n}$ for $t \in \ooint{S_n,S_{n+1}}$}\\
  \mbox{\textcolor{white}{Set}  $X_{S_{n+1}} = X_{S_n} + T_{n+1} Y_{S_n}$}\\
\eIf{$T_{n+1} = T_{n+1}^{\text{Ev}}$}{ \mbox {Set $Y_{n+1} \sim \kernelQ((X_{S_{n+1}},Y_{S_n}),\cdot)$ and Set $T^{\text{Ref}}_{n+2} = T^{\text{Ref}}_{n+1} - T_{n+1}$}}
   {Set $Y_{(n+1)\rmT} \sim \loiy$ and  $T^{\text{Ref}}_{n+2} = T$}
    }
\end{algorithm}
\newpage

\subsection{Factorized Piecewise Deterministic Monte Carlo methods}
\label{sec:DPDMC}

When the potential $U$ can be written as a sum of terms, $ U(x) =
\sum_{i=1}^d U_i(x)$ or considering directly the decomposition of the
gradient $\nabla U$ over the direction, it can be convenient to
exploit this decomposition through the implementation of the
factorized Metropolis filter \citep{michel:kapfer:krauth:2014}, for
example to exploit some symmetries of the problem or reduce the complexity 
\citep{michel:2017}. It finds its equivalent in PDMC by
considering a superposition of Poisson processes
\citep{peters:dewith:2012,michel:kapfer:krauth:2014,bouchard:vollmer:doucet:2016}. The results developed in \Cref{sec:derivation-new-pdmps} can be
generalized using this property, as we explain in more details in the
supplement \Cref{sec:DPDMC}.

We consider in this section the following decomposition of the potential $U$,
\begin{equation}
  \label{eq:decomp_U}
  U(x) = \sum_{i=1}^N U_i(x) \eqsp,
\end{equation}
where $N \in \nset^*$ and for any $i \in \{1,\ldots,N\}$,
$U_i : \rset^d \to \rset$ are continuously differentiable
function. Similarly to \cite{michel:kapfer:krauth:2014} and
\cite[Section 3]{bouchard:vollmer:doucet:2016}, we can adapt the
construction of PDMCs which target $\pi$ described in
\Cref{sec:derivation-new-pdmps} to exploit this decomposition. Indeed,
it can be more convenient to compute event times associated with the rates
\begin{equation}
  \label{eq:lambda_i}
  \lambda_i(x,y) = \ps{y}{\nabla U_i(x)}_+, \text{ for any } 
(x,y) \in \rset^d \times \msy \eqsp, \, i \in \{1,\ldots,N\} \eqsp, 
\end{equation}
rather than to consider the rate given in \eqref{eq:definition_rate}
directly.  To do so, we need to introduce $N$ Markov kernel
$(Q_i)_{i \in \{1,\ldots,N\}}$ on $(\rset^{2d},\mcb{\rset^d})$ such that
  for any $i \in \{1,\ldots,N\}$, $Q_i$ satisfies the conditions of
  \Cref{theo:invariance_BPS_generalized} relatively to $U_i$ and
  therefore \eqref{eq:extended_global_balance} is satisfied with
  respect to $U_i$, \ie~for any $f \in \rmC^1_c(\rset^{2d})$, for
  almost all $x \in \rset^d$,
  \begin{equation}
\label{eq:extended_global_balance_i}
  \int_{\rset^d} \int_{\rset^d} \ps{y}{\nabla U_i(x)}_+ f(x,\ty) \kernelQ_i((x,y),\rmd \ty)  \rmd \loiy(y) =  \int_{\rset^d} \ps{\ty}{\nabla U_i(x)}_- f(x,\ty)  \rmd \loiy(\ty) \eqsp.
\end{equation}

Consider now the PDMP with characteristic $(\varphi,\lambdatot,\Qtot)$ where $\varphi$ is defined by \eqref{eq:differential_flow},
\begin{equation*}
  \lambdatot = \sum_{i=1}^N \lambda_i \eqsp, \qquad \Qtot = \lambdatot^{-1}\sum_{i=1}^N \lambda_i Q_i \eqsp. 
\end{equation*}
Its generator is given for any $f \in \rmC^1(\rset^{2d})$ and $(x,y) \in \rset^{2d}$ by
\begin{equation*}
  \generatortot f(x,y) = \ps{y}{\nabla_xf(x,y)} + \sum_{i=1}^N \ps{y}{\nabla U_i(x)}_+ \defEns{\int_{\rset^d} f(x,\ty) Q_i((x,y),\rmd \ty) - f(x,y)} \eqsp.  
\end{equation*}
Using that \eqref{eq:extended_global_balance_i} is satisfied and the
same reasoning as for the proof of
\Cref{theo:invariance_BPS_generalized}, we get that
$\int_{\rset^{d} \times \msy} \generatortot f(x,y) \rmd \tpi(x,y) = 0$
and $\tilde{\pi}$ given by \eqref{eq:extended_target} is invariant. A
procedure to sample such a PDMP is given in  \Cref{algo:Fact_Forward_MCMC}. We refer to
\cite[Section 3.3]{bouchard:vollmer:doucet:2016} for more algorithmic considerations.

\begin{algorithm}[h]
 \caption{ \label{algo:Fact_Forward_MCMC} Factorized PDMP process}
  \KwData{A potential $U$ satisfying the decomposition \eqref{eq:decomp_U},  Markov kernels $(Q_i)_{i \in \{1,\ldots,N\}}$ on $(\rset^{2d},\mcb{\rset^d})$ satisfying \eqref{eq:extended_global_balance_i}  and initial points $(X_0,Y_0)$}
 \KwResult{Factorized PDMC  $(X_t,Y_t)_{t \geq 0}$}
 Initialize $S_0 =0$ and a sequence of \iid~exponential random variables $(E_{i,j})_{\substack{i \in \{1,\ldots,N\},\, j \geq 1}}$ with parameter $1$\\
 \For{$n \geq 0$}{
   \For{$i \in \{1,\ldots,N\}$}{
        \mbox{Set $T_{n+1}^{\text{Ev},i} = \inf\{ t \geq 0 \, : \, \int_{0}^t \ps{Y_{S_n}}{\nabla U_i(X_{S_n} + s Y_{S_n})}_+ \rmd s \geq E^i_{n+1}\}$}
     }
   Set $I = \argmin_{i \in \{1,\ldots,N\}} T_{n+1}^{\text{Ev}},i$, $T_{n+1} = T_{n+1}^I$ and   $S_{n+1} = S_n + T_{n+1}$\\
     \mbox{Set   $Y_t = Y_{S_n}$,  $X_{t} = X_{S_n} + (t-S_n) Y_{S_n}$ for $t \in \ooint{S_n,S_{n+1}}$}\\\mbox{Set $X_{S_{n+1}} = X_{S_n} + T_{n+1} Y_{S_n}$}\\
     Sample $Y_{S_{n+1}} \sim  Q_I((X_{S_{n+1}}, Y_{S_n}),\cdot)$.
   }
      \end{algorithm}

      \begin{example}[Logistic regression]
        \label{ex:log_ref_event_times}
        In the Bayesian logistic regression problem, the potential $U$
        given by \eqref{eq:U_log_reg} can be decomposed for
        $\theta \in \rset^d$ as $U(\theta) = \sum_{i=0}^N U_i(\theta)$
        where $U_0(\theta) = \norm[2]{\theta}/(2\varsigma^2)$ and for
        $i \in \{1,\ldots,N\}$,
        \begin{equation}
          \label{eq:def_u_i_log_reg}
          U_i(\theta) = -y_i\ps{x_i}{\theta} +
        \log\parentheseDeux{1+\exp(\ps{x_i}{\theta})} = \log\parentheseDeux{ \frac{1+\exp(\ps{x_i}{\theta})}{\exp(y_i\ps{x_i}{\theta})}} \eqsp.
        \end{equation}
        The event
        times associated with $U_0$ can be sampled from the procedure
        described in \ref{sec:sampling-event-times}. On the other hand,  the
        events for $U_i$, $i \in \{1,\ldots,N\}$, can be computed as
        follow. First note since for any $a>0$,  $t \mapsto 1+a \rme^{t}$ is non-decreasing on $\rset$ and $t \mapsto (1+a \rme^{t})/(a \rme^{t})$  is non-increasing on $\rset$, we get  by definition \eqref{eq:def_u_i_log_reg} that for any $i\in\{1,\ldots,N\}$,
        $\theta,v \in \rset^d$,
        \begin{equation}
          \label{eq:sign_nabla_u_i_log_reg}
        \begin{aligned}
                     \text{ if $y_1=0$ },
           \begin{cases}
             s \mapsto U_i(\theta + sv ) \text{ is non-decreasing } & \text{ if $\ps{x_i}{v} \geq 0$} \\
                          s \mapsto U_i(\theta + sv ) \text{ is non-increasing } & \text{ if $\ps{x_i}{v} <  0$} 
                        \end{cases}
                        \eqsp, \\ 
           \text{ if $y_1=1$ },
           \begin{cases}
                          s \mapsto U_i(\theta + sv ) \text{ is non-increasing } & \text{ if $\ps{x_i}{v} \geq 0$} \\
                          s \mapsto U_i(\theta + sv ) \text{ is non-decreasing } & \text{ if $\ps{x_i}{v} <  0$}  \eqsp.
           \end{cases}
         \end{aligned}
       \end{equation}
In addition, for any $i\in\{1,\ldots,N\}$,
        $\theta,v \in \rset^d$, $s \in \rset$,
        \begin{equation}
          \label{eq:nabla_u_i_log_reg}
          \ps{v}{\nabla U_i(\theta+sv) } = \ps{v}{x_i} \defEns{-y_i  + \parenthese{1+\rme^{-s \ps{v}{x_i} - \ps{\theta}{x_i}}}^{-1}} \eqsp.
        \end{equation}
        Then, given the exponential random $E$ and a current position $\theta$ and direction $v$, the calculation of the event time $T_i$ can be decomposed in two steps. First check that $s \mapsto U_i(\theta + sv )$ is increasing, which is equivalent by \eqref{eq:sign_nabla_u_i_log_reg} and \eqref{eq:nabla_u_i_log_reg} to check that $(y_i-1/2)\ps{x_i}{\theta} < 0$, otherwise set $T_i = \plusinfty$. If $s \mapsto U_i(\theta + sv )$ is increasing, then   compute $T_i = \inf\{ s \geq 0 \, : \,  U_i(x+sv) - U_i(x) \geq E \}$, which is equivalent by continuity to solve the equation
        $ U_i(x+T_iv) - U_i(x) = E$. By \eqref{eq:def_u_i_log_reg}, this equation is equivalent to solve, setting $b= \ps{v}{x_i}$,
        \begin{align*}
      \log(1+\rme^{\ps{\theta}{x_i}}\rme^{bT_i})-\log(1+\rme^{\ps{\theta}{x_i}}) - y_i(bT_i)=E  \eqsp. \\
        \end{align*}
Therefore, the equation        $ U_i(x+T_iv) - U_i(x) = E$         has a unique solution given by
\begin{align}
  \label{eq:def_t_i_log_ref_0}
   T_i =\left. \log\parentheseDeux{\rme^E + ( \rme^E -1 ) ((1-y_i)\rme^{-\ps{\theta}{x_i}} + y_i\rme^{\ps{\theta}{x_i}}  )} \middle/|\ps{v}{x_i}| \right. 
        \end{align}
         The pseudo-code associated with the computation of $T_i$ is given in \Cref{algo:compute_T_i_log_reg}.    
\begin{algorithm}[h!]
 \caption{ \label{algo:compute_T_i_log_reg} Event time computation for Bayesian logistic regression problem}
 \KwData{A potential $U_i$ of the form \eqref{eq:def_u_i_log_reg}, for $x_i \in \rset^d$ and $y_i \in \{0,1\}$, $\theta \in \rset^d$, $v \in \rset^d$ and $E \geq 0$}
 \KwResult{Solution  $T_i \in \rset_+ \cup \{\plusinfty\}$ of $U_i(x+T_iv) - U_i(x) \geq E$}
 \eIf{$(y_i-1/2)\ps{x_i}{\theta} >0$}
 {
   Set $T_i= \plusinfty$ \hfill\mbox{\emph{No solution}}
 }
 {
   Set $T_i$ according to \eqref{eq:def_t_i_log_ref_0}
 }
\end{algorithm}
      \end{example}

\subsection{Sampling of event times for $d$-dimensional Gaussian target distributions}
\label{sec:sampling-event-times}

We consider in this section a zero-mean Gaussian distribution of
dimension $d$ associated to a covariance matrix $\Sigma$ and described
by the potential $U$, given for any $x \in \rset^d$ by
$U(x) = \ps{x}{\Sigma^{-1} x}/2$, and a PDMP whose event times are
associated to the rate given in \eqref{eq:definition_rate}.  Starting
from $(x_0, y_0)$, we want to compute the next event time $T$, defined
through the equation $\int_{0}^t \lambda(x_0 + ty_0,y_0) \rmd t = V$,
where $V$ is a exponential random variable with parameter $1$, leading
to
 \begin{equation}
  \label{eq:int_poisson}
  \int_{0}^T \max\left(0,2u(t + T_0)\right) \rmd t = V,
\end{equation}
with $u = \ps{y_0}{\Sigma^{-1} y_0} /2$ and $T_0 = (\ps{y_0}{\Sigma^{-1} x_0}/(2U_0)$ , which gives,
\begin{equation}
  \label{eq:T_ev_Gauss}
\left\{\begin{array}{ll}
   T = - T_0 + \sqrt{V/u + T_0 ^2} & \text{if } T_0 \geq 0\\
   T = - T_0 + \sqrt{V/Yu} & \text{otherwise}.
\end{array}\right.
\end{equation}

In a general manner, solving integral as \eqref{eq:int_poisson} comes
down to locating the zeros of the unidimensional function $
\ps{y}{\nabla U(x+ty)}$ and summing up the positive intervals.

\subsection{Sampling of event times for a Poisson-Gaussian Markov random field}
\label{sec:sampling-poisson}

We consider in this section a Poisson-Gaussian Markov random field
model as described in \Cref{sec:poisson-gauss}, where the observations
$\mathbf{Y} = (Y_{i,j})_{(i,j) \in \{1,\ldots,\tilde{d}\}^2} \in
\nset^{\tilde{d}^2} $
are supposed to be independent samples such that for any
$(i,j) \in \{1,\ldots,\tilde{d}\}^2$, $Y_{i,j}$ has a Poisson
distribution with parameter $\exp(x_{i,j})$ with $x \in \rset$. The
parameter $\bfx = (x_{i,j})_{(i,j) \in \{1,\ldots,\tilde{d}\}^2}$ is
assumed to be a Gaussian random field, \ie~the prior distribution is
set to be the zero-mean Gaussian distribution with covariance matrix
$\tilde{\Sigma}_{(i,j),(\tilde{i},\tilde{j})}$. The corresponding
function rate $\lambda$ \eqref{eq:definition_rate} is given for any
$x,v \in \rset^d$ by
\begin{equation}
  \rate(x,v)\!=\!\left\langle \sum_{i,j,\tilde{i},\tilde{j}=1}^{\tilde{d}} v_{i,j}\tilde{\Sigma}_{(i,j),(\tilde{i},\tilde{j})}^{-1}x_{\tilde{i},\tilde{j}} + \sum_{i,j=1}^{\tilde{d}}\{\exp(x_{i,j})v_{i,j} -v_{i,j} Y_{i,j}\}\right\rangle_+\eqsp,
\end{equation}
which is bounded by $\lambda_B(x,y) = \lambda_{B,G}(x,v) +
\sum_{i,j=1}^{\tilde{d}} \lambda_{B,i,j}(x,v)$ with,
\begin{equation}
  \lambda_{B,G}(x,v) = \left\langle \sum_{i,j,\tilde{i},\tilde{j}=1}^{\tilde{d}} v_{i,j} \left(\tilde{\Sigma}_{(i,j),(\tilde{i},\tilde{j})}^{-1}x_{\tilde{i},\tilde{j}}- Y_{i,j}\right)\right\rangle_+  \eqsp,  
  \lambda_{B,i,j}(x,v) =\exp(x_{i,j})\langle v_{i,j}\rangle_+ \eqsp.
\end{equation}
Starting from
$(x_0,v_0)\in \rset^{\tilde{d}^2}\times \rset^{\tilde{d}^2}$, we then
sample by thinning the next event time $T$ as follows:
\begin{enumerate}
\item We compute the next event time $T_B$ associated with $\lambda_B$
  directly:
\begin{itemize}
\item The event times $T_G$ associated wih $\lambda_{B,G}$ can be
  sampled from the procedure described in
  \Cref{sec:sampling-event-times}, with
  $u=\sum_{i,j,\tilde{i},\tilde{j}=1}v_{0,i,j}\tilde{\Sigma}_{(i,j),(\tilde{i},\tilde{j})}^{-1}v_{0,\tilde{i},\tilde{j}}$
  and
  $T_0= \sum_{i,j,\tilde{i},\tilde{j}=1}^{\tilde{d}}
  v_{0,i,j}(\tilde{\Sigma}_{(i,j),(\tilde{i},\tilde{j})}^{-1}x_{0,\tilde{i},\tilde{j}}
  -Y_{i,j})/(2u)$.
\item Concerning the processes with rates $\lambda_{B,i,j}$, a finite
  event time $T_{i,j}$ exists only if $v_{i,j}>0$ and is then given by
  $\int_0^{T_{i,j}}\exp(x_{0,i,j} + tv_{0,i,j})v_{0,i,j}dt= V$, where
  $V$ is a exponential random variable with parameter $1$. It yields
  $T_{i,j} = \log(1+V\exp(-x_{0,i,j}))/v_{0,i,j}$.
\item Eventually, the next event time of the process of rate
  $\lambda_B$ is $T_B = \min(T_G,\{T_{i,j}; v_{i,j}>0\})$.
\end{itemize}
\item The next event time $T$ associated to $\lambda$ is set to $T_B$
  with probability $\lambda(x_B,v_B)/\lambda_B(x_B,v_B)$, with
  $(x_B, v_B)=(x_0+T_Bv_0,v_0)$.
\item If it is rejected, the procedure is applied since step $1$ but
  starting from $(x_B, v_B)$, until acceptance.
\end{enumerate}

\section{Addition to the numerical experiments}

We first display a comparaison of PDMC-type schemes and a
Hastings-Metropolis algorithm on the anisotropic zero-mean Gaussian
distribution in \Cref{sec:numer-metro}.

We then motivate in this section the choice of $\Kp$ and $\Kn$ used
for the numerical experiments. We first display the performances of
the different choices of $\Kp$ (direct sampling or Metropolis-based
sampling) in \Cref{sec:numer-comp-betw} and then compare different
  orthogonal refreshment and standard refreshment schemes in
  \Cref{sec:numer-comp-diff}.

  Finally, we give some details on the role played by the choice of
  $\rmT$ for the refreshment time in \Cref{sec:rate} for the
  anisotropic zero-mean Gaussian distribution in \Cref{sec:gaussian}
  and we display the ESS obtained at small dimensions for the mixture
  of Gaussian distribution of \Cref{sec:mixgaussian}.

\subsection{Numerical comparisons between PDMC methods and a Hastings-Metropolis algorithm}
\label{sec:numer-metro}

Considering the anisotropic zero-mean Gaussian distribution with
covariance matrix given by (\ref{eq:Gaussian}), a Hastings-Metropolis
algorithm is tuned in order to maximise the decorrelation observed on
the observables (the potential $U$, $\norm{x}$, $x$) for the range of
considered dimensions ($25, 50, 100, 200,400$). Successives proposed
moves correspond to an update along a random vector of uniform
direction in the hypersphere and of uniform norm between $0$ and
$\delta=10$.

\begin{figure}
\begin{center}
\includegraphics[width=1.0\textwidth]{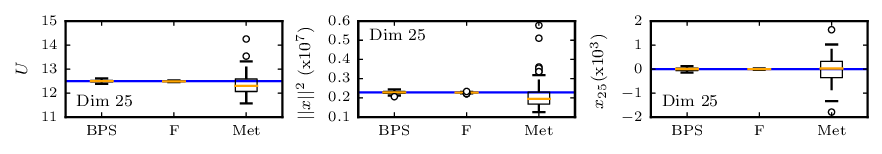}\\
\includegraphics[width=1.0\textwidth]{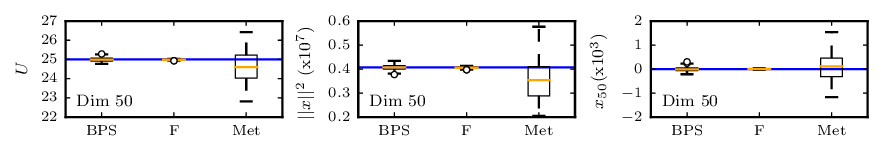}\\
\includegraphics[width=1.0\textwidth]{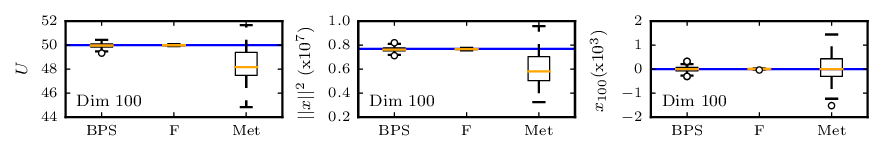}\\
\includegraphics[width=1.0\textwidth]{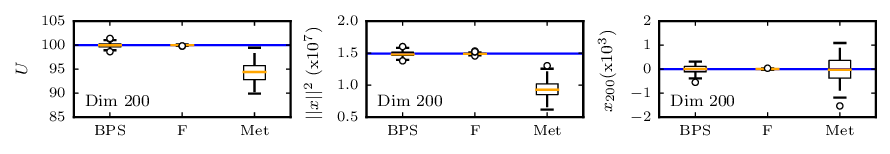}\\
\includegraphics[width=1.0\textwidth]{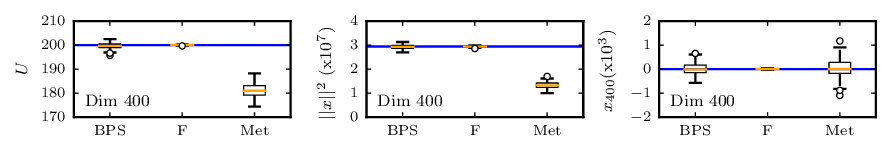}\\
\caption{\label{fig:Gauss_Num_Metro} Box plots of $U$, $||x||^2$ and
  $x$ for the ill-conditioned zero-mean Gaussian distribution with
  covariance matrix given by \eqref{eq:Gaussian} and $d=25,50,100,200$
  and $400$. Each box represent the results of 100 runs of $10^5$
  samples separated by a fixed CPU time $\sim 7. 10^{-3}s$,
  resulting in $200$ updates for the Hastings-Metropolis (Met)
  algorithm and $\sim 55$ events for the BPS and Forward Ref (F).}
\end{center}
\end{figure}

\subsection{Numerical comparisons between direct and Metropolis Forward event-chain methods}
\label{sec:numer-comp-betw}

We compare the performances given by the choice of $\Kp$ between the
direct, independent Metropolis and random-walk Metropolis ($\delta \in
[-0.5,0.5]$) schemes, with a refreshment set to an orthogonal switch
at fixed time $\rmT$ and $\lambdab=0$, corresponding to
\Cref{algo:update_Ext_Forward_MCMC}.

\begin{figure}
\begin{center}
\includegraphics[width=1.0\textwidth]{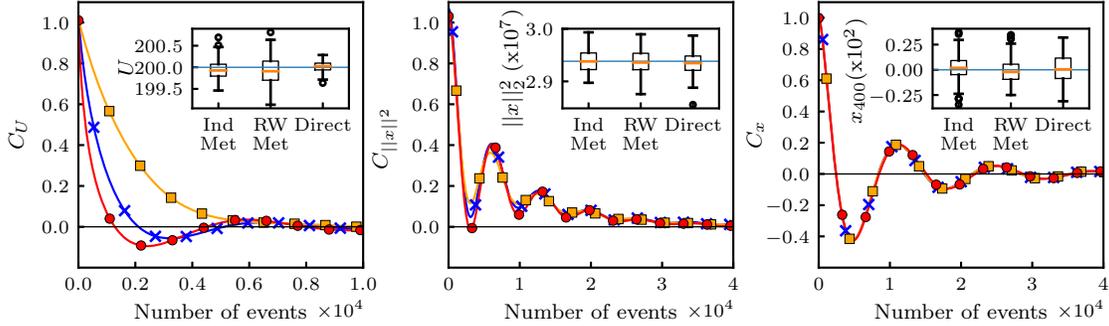}
 \caption{Autocorrelation functions $C$ of the potential $U$
   (\textbf{Left}), $||x||^2$ (\textbf{Middle}) and $x$
   (\textbf{Right}) for the ill-conditioned zero-mean Gaussian
   distribution with covariance matrix given by (\ref{eq:Gaussian})
   and $d=400$ for Forward EC schemes with independent-Metropolis
   sampling (blue, crosses), random-walk Metropolis sampling (yellow,
   squares) and direct sampling (red, circles). Insets correspond to
   the respective boxplots for 100 runs of $10^5$ samples separated by
   a fixed time $T=500$ corresponding to an average of $55$ events. \label{fig:Gauss_Num_Pick}}
\end{center}
\end{figure}

For the anisotropic Gaussian distribution of \Cref{sec:gaussian}, we
set $\rmT=500$ ($\sim 55$ events in average).  The autocorrelation
functions for the potential $U$, the squared norm $\norm{x}^2$ and $x$
are represented on Fig. \ref{fig:Gauss_Num_Pick} for $d=400$, as long
as the respective boxplots. All schemes show similar decorrelation
behavior for $x$, but for the potential and the norm, the sampling
scheme mixing the less (random-Walk Metropolis) is the slowest and the
sampling scheme mixing the most (direct) is the fastest. The direct
sampling scheme is also more efficient regarding the norm. 

For the mixture of Gaussian distributions considered in
\Cref{sec:mixgaussian}, \Cref{tab:Ran_ESS} summarizes the ESS for
$x$ and $\norm{x}^2$ for $d=2,4$ and $5$. At very small dimension
$d=2$, less-mixing schemes based on Metropolis updates appears to be
slightly faster but quickly, as dimension increases, results are
similar.

\subsection{Numerical comparisons for different refreshment strategies
  for the direct Forward event-chain method}
\label{sec:numer-comp-diff}

\begin{figure}
\begin{center}
\includegraphics[width=1.0\textwidth]{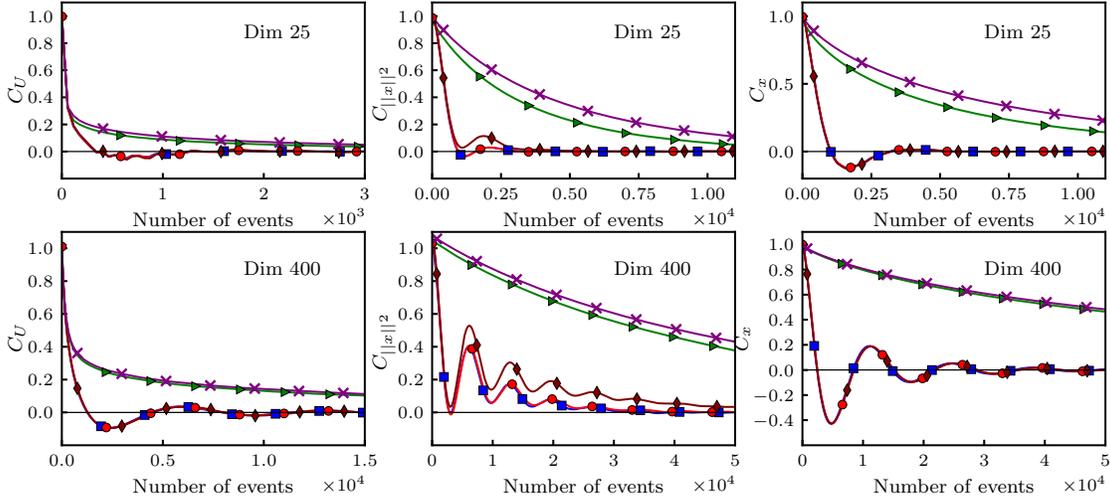}
\caption{Autocorrelation functions $C$ of the potential $U$
  (\textbf{Left}), the squared norm $||x||^2$ (\textbf{Middle}) and
  $x$ (\textbf{Right}) for the ill-conditioned zero-mean Gaussian
  distribution with covariance matrix given by \eqref{eq:Gaussian} and
  $d=25$ (\textbf{Top}) and $d=400$ (\textbf{Bottom}) for direct
  Forward EC scheme with an orthogonal switch (red, circle), a perp
  orthogonal switch (maroon, slim diamond), a ran-$2$-p orthogonal switch (blue, square), a full orthogonal refresh
  (green, right triangle) and a full refresh (purple, cross). Every
  $\Kn$ is set to its positive type. \label{fig:Gauss_Num_Ref}}
\end{center}
\end{figure}

Fixing $\Kp$ to the direct-sampling kernel, we now compare different
refreshment choices at fixed $\rmT$ for $\Kn$ on the experiment with
the anisotropic Gaussian distribution as described by
\eqref{eq:Gaussian}, as illustrated by Fig. \ref{fig:Gauss_Num_Ref},
where the autocorrelation functions for $U, ||x||^2$ and $x$ are
shown. Refreshment schemes are then separated into two groups
depending on their action and the obtained decorrelation:
\begin{itemize}
\item the sparse-orthogonal group, where the refreshment only
acts on a few orthogonal components, as 
 \begin{itemize} 
\item Orthogonal switch: two orthonormal vectors $e_1,e_2$ of the orthogonal plane are defined by the Gram-Schmidt process and $y$ is transformed into $\tilde{y} = y + (\ps{e_2}{y}-\ps{e_1}{y})(e_1 -e_2)$
\item Perpendicular orthogonal switch: same as above but $y$ is set to $\tilde{y} = y-\ps{e_1}{y}(e_1-e_2) -\ps{e_2}{y}(e_1 + e_2)$.
\item ran-$2$-orthogonal: random rotation of $2$ orthogonal
  components defined by the Gram-Schmidt process.
\end{itemize}
\item the global group, where all components can be
resampled, as 
\begin{itemize}
\item Full-orthogonal refresh: full refreshment of the orthogonal components of the direction. 
\item Full refresh: full refreshment of the direction.
\end{itemize}
\end{itemize}
More details on the definition of $\Kp$ can be found in Section \ref{sec:choice_ortho}.

\begin{figure}
\begin{center}
\includegraphics[width=1.0\textwidth]{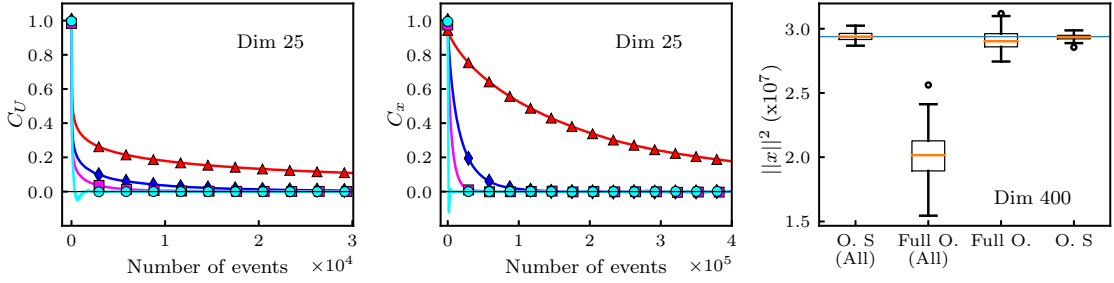}
\caption{Autocorrelation functions $C$ of the potential $U$
  (\textbf{Left}) and $x$ (\textbf{Middle}) for $d=25$ and box plots
  for $\norm{x}^2$ (\textbf{Right}) for $d=400$ for the
    ill-conditioned $d-$dimensional zero-mean Gaussian distribution
    with covariance matrix given by \eqref{eq:Gaussian} for direct
    Forward EC scheme to an orthogonal switch (O. S.) at fixed time (cyan,
    circle) and at all events (blue, diamond) and a full-orthogonal
    refresh (Full O.) at fixed time (magenta, square) and at all events (red,
    diamond). Every $\Kn$ is set to its positive type. Box plots are
    based on $100$ runs of $10^5$ samples separated by $\rmT=500$. \label{fig:Gauss_Num_Ref_SparseFull}}
\end{center}
\end{figure}

For the small
dimension ($d=25$) as the bigger one ($d=400$), a random-walk behavior
appears in the global group, whereas in the sparse-orthogonal group, the
antithetic correlations given by the ballistic trajectories are
preserved and a faster decorrelation is achieved. In this group, the
orthogonal switch scheme is the most efficient. In the global group,
we observe that updating all the components but the one parallel to
the gradient leads to a small acceleration in comparison to the
standard full refreshment scheme. Finally,
\Cref{fig:Gauss_Num_Ref_SparseFull} compares an orthogonal switch
and a full-orthogonal refresh set at a fixed time $\rmT$ and at all
event (no tuning of $\rmT$). It appears clearly that the
orthogonal switch remains competitive without any tuning of $\rmT$,
whereas the full-orthogonal refresh shows some convergence issue in
that situation.

\begin{figure}
\begin{center}
\includegraphics[width=0.8\textwidth]{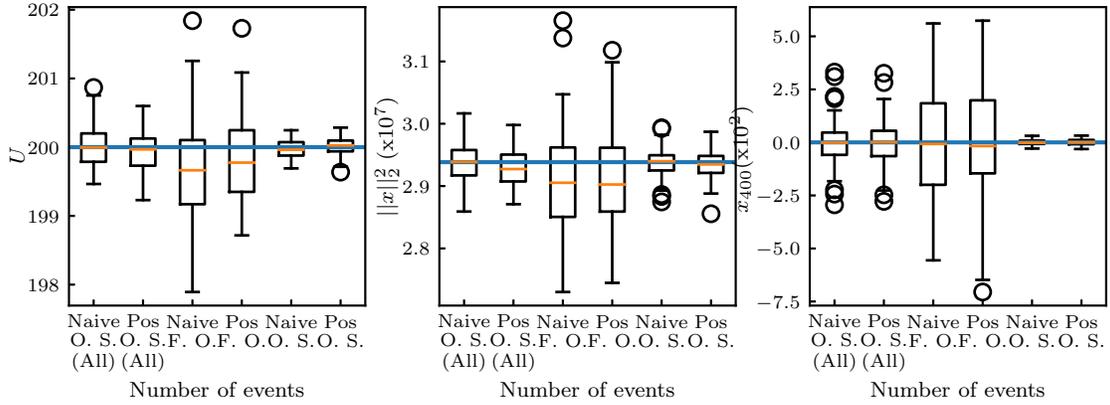}
\caption{\label{fig:Gauss_Num_NaivePos} Box plots of $U$, $||x||^2$ and $x$ for the ill-conditioned
  zero-mean Gaussian distribution with covariance matrix given by
  \eqref{eq:Gaussian} and $d=400$. Each box represent the results of
  100 runs of $10^5$ samples separated by a fixed time $\rmT=500$ ($\sim 55$ events). Pos stands for Positive, Orth S. for orthogonal switch and F. O. for full-orthogonal refresh.}
\end{center}
\end{figure}

In Figure \ref{fig:Gauss_Num_Ref} and
\Cref{fig:Gauss_Num_Ref_SparseFull}, only positive-type $\Kn$ are
exhibited. Figure \ref{fig:Gauss_Num_NaivePos} compares positive to
naive schemes and shows that the positive schemes are slightly better
for the sampling of the anisotropic distribution.

\begin{figure}
\begin{center}
\includegraphics[width=0.8\textwidth]{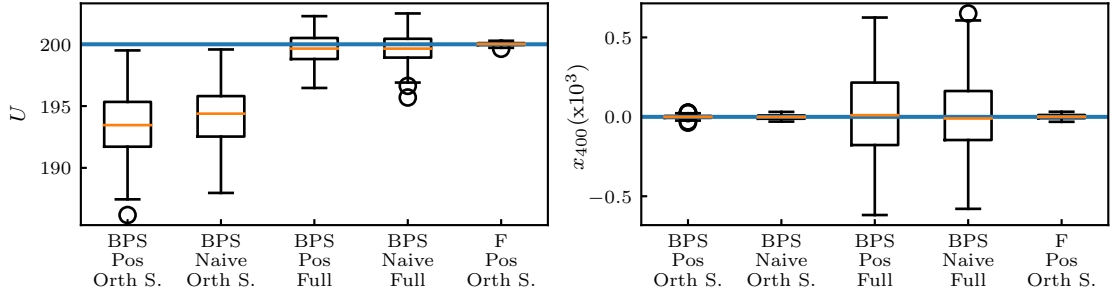}
\caption{\label{fig:Gauss_Num_BPSOrth} Box plots of $U$ and $||x||^2$ for the ill-conditioned
  zero-mean Gaussian distribution with covariance matrix given by
  \eqref{eq:Gaussian} and $d=400$. Each box represent the results of
  100 runs of $10^5$ samples separated by a fixed time $\rmT=500$ ($\sim 55$ events). Pos stands for Positive, Orth S. for orthogonal switch and Full for Full Ref.}
\end{center}
\end{figure}

Finally, we show in Figure~\ref{fig:Gauss_Num_BPSOrth} the performance
of BPS with different refreshment schemes, as to check whether
accelerations can be achieved only by a choice of $\Kn$ different from
$\Id$, while still keeping $\Kp$ set to the deterministic choice of
the reflection. We consider the standard Full Ref in a naive and
positive implementation and a orthogonal switch at fixed time $\rmT$
associated with $\lambdab=0$, also in a positive and naive
settings. Positive and naive types give similar results. However, it
appears clearly that BPS set to the sparse-orthogonal refreshment
scheme of the orthogonal switch is not able to recover a correct
estimate of the potential $U$, while the decorrelation in respect of
$x$ is fast. BPS requires indeed a strong refreshment as the
deterministic choice of $\Kp$ leads to poor mixing for $U$, but at the cost of a slow decay of $x$.

\subsection{Impact of the choice of the refreshment parameters}\label{sec:rate}

\begin{figure}
\begin{center}
\includegraphics[width=1.0\textwidth]{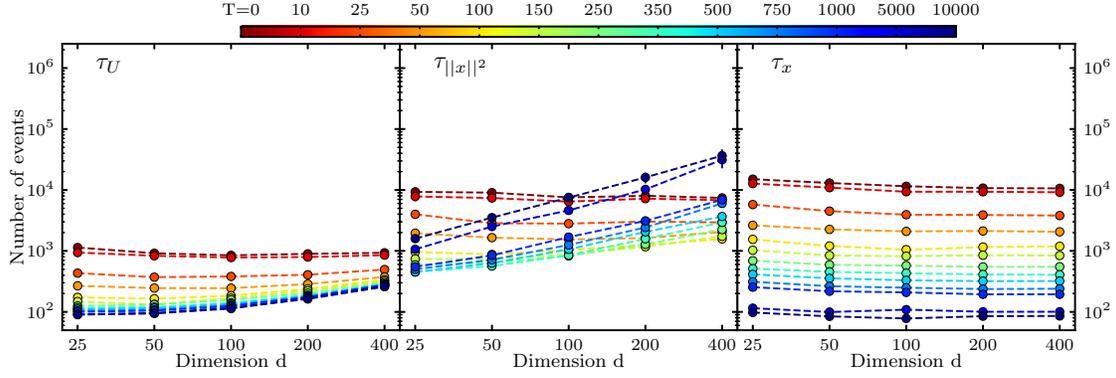}
\caption{Integrated autocorrelation times $\tau$ of $U$
  (\textbf{Left}), $\norm{x}^2$ (\textbf{Middle}) and $x$
  (\textbf{Right}) for the anisotropic Gaussian distribution for Forward Ref for different refreshment time $\rmT$ ($\rmT=0$ corresponds to Forward Ref All). Error bars may be covered by the markers. \label{fig:Gauss_Num_IntForward}}
\end{center}
\end{figure}

\begin{figure}
\begin{center}
\includegraphics[width=1.0\textwidth]{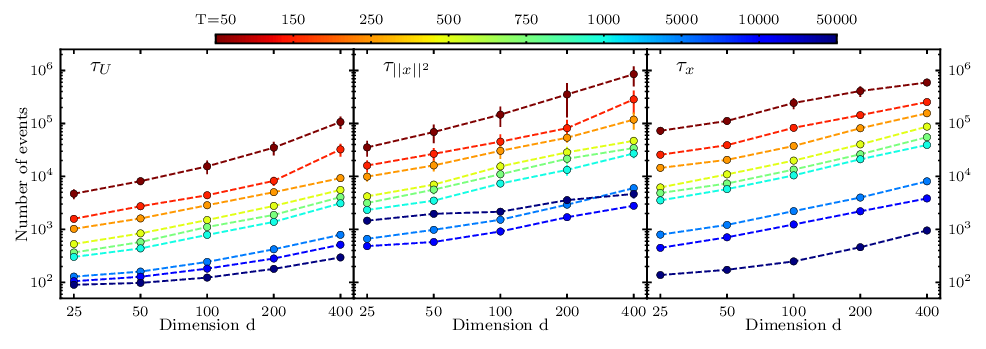}
\caption{Integrated autocorrelation times $\tau$ of $U$
  (\textbf{Left}), $\norm{x}^2$ (\textbf{Middle}) and $x$
  (\textbf{Right}) for the anisotropic Gaussian distribution for
  Forward Full Ref for different refreshment time $\rmT$. Error bars
  may be covered by the markers. \label{fig:Gauss_Num_IntRefs}}
\end{center}
\end{figure}

\begin{figure}
\begin{center}
\includegraphics[width=1.0\textwidth]{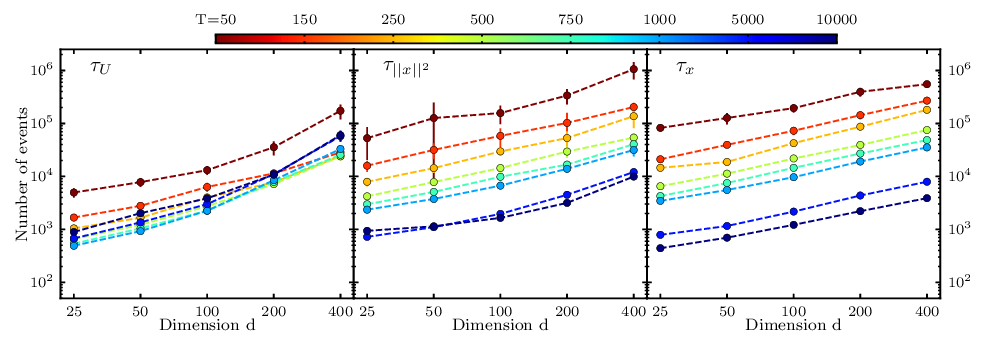}
\caption{Integrated autocorrelation times $\tau$ of $U$
  (\textbf{Left}), $\norm{x}^2$ (\textbf{Middle}) and $x$
  (\textbf{Right}) for the anisotropic Gaussian distribution for BPS Full Ref for different refreshment time $\rmT$. Error bars may be covered by the markers. \label{fig:Gauss_Num_IntBPS}}
\end{center}
\end{figure}

\begin{figure}
\begin{center}
\includegraphics[width=1.0\textwidth]{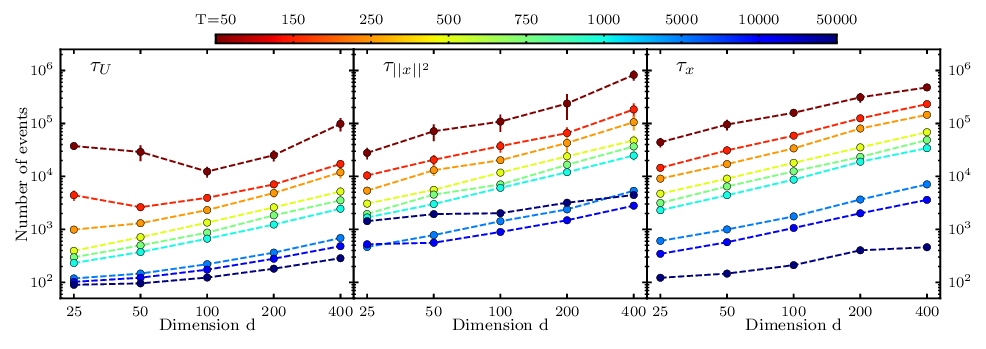}
\caption{Integrated autocorrelation times $\tau$ of $U$
  (\textbf{Left}), $\norm{x}^2$ (\textbf{Middle}) and $x$
  (\textbf{Right}) for the anisotropic Gaussian distribution for
  direct Forward EC set to full-orthogonal refresh for different
  refreshment time $\rmT$. Error bars may be covered by the markers.}
\label{fig:Gauss_Num_IntGaussRef}
\end{center}
\end{figure}

We consider the zero-mean Gaussian distribution with covariance matrix
given by (\ref{eq:Gaussian}) and study the effects of the refreshment
time $\rmT$ tuning on the integrated autocorrelation times for the
Forward Ref (see Figure~\ref{fig:Gauss_Num_IntForward}), Forward Full
Ref (see Figure~\ref{fig:Gauss_Num_IntRefs}), BPS Full Ref (see
Figure~\ref{fig:Gauss_Num_IntBPS}) and Forward Ref set to a
full-orthogonal refreshment (see \Cref{fig:Gauss_Num_IntGaussRef}).

\begin{figure}
\begin{center}
\includegraphics[width=1.0\textwidth]{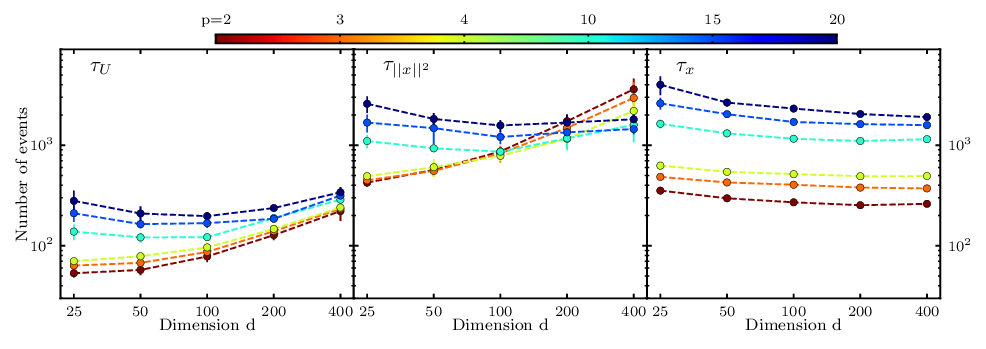}
\caption{Integrated autocorrelation times $\tau$ of $U$
  (\textbf{Left}), $\norm{x}^2$ (\textbf{Middle}) and $x$
  (\textbf{Right}) for the anisotropic Gaussian distribution for
  direct Forward EC set to ran-p-orthogonal refresh for different
  value of $p$. Error bars may be covered by the markers.}
\label{fig:Gauss_Num_IntDim}
\end{center}
\end{figure}

\begin{figure}
\begin{center}
\includegraphics[width=1.0\textwidth]{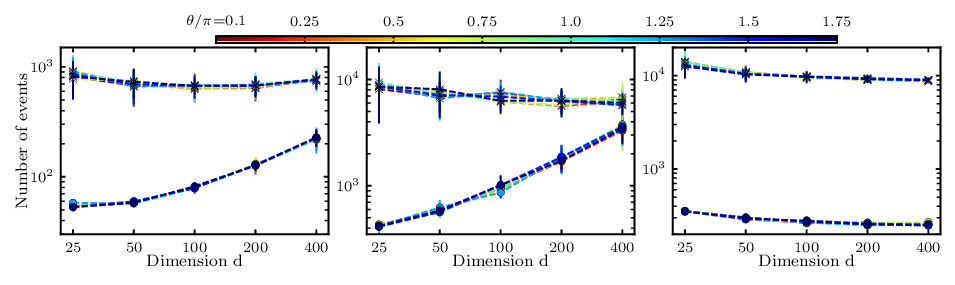}
\caption{Integrated autocorrelation times $\tau$ of $U$
  (\textbf{Left}), $\norm{x}^2$ (\textbf{Middle}) and $x$
  (\textbf{Right}) for the anisotropic Gaussian distribution for
  direct Forward EC set to $2$-orthogonal refresh for different
  value of $\theta$ and $\rmT=0$ (crosses) and $\rmT=500$ (circles). Error bars may be covered by the markers.}
\label{fig:Gauss_Num_IntTheta}
\end{center}
\end{figure}

A first observation is that the scaling with the dimension of the
integrated autocorrelation time of $x$ is similar for any choice of
$\rmT$ and the offset decreases as $\rmT$ increases (less
randomization), for all schemes. For the potential $U$ and the squared
norm $\norm{x}^2$, on the contrary, there is a trade-off to find
between controlling the random-walk behavior and trapping the
process into a loop. Forward Ref appears as the most robust concerning
this tuning, as all choices of $\rmT$ are in the same range. BPS Full
Ref, on the opposite, needs $\rmT$ small enough to decorrelate the potential
$U$, at the cost of the norm decorrelation. Comparing it to the
results for Forward Full Ref, we can observe that the
direct-sampling scheme helps with the decorrelation of $U$ and allows
to set $\rmT$ to very high values. The choice $\rmT=10^4$ appears as
an optimal, leading to a maximal integrated time of order $2\times
10^3$, which is competitive with Forward Ref. However, Forward Full
Ref is more sensitive to the tuning of $\rmT$ than Forward Ref. Same behavior can be observed for Forward Ref with full-orthogonal refreshment, as displayed in \Cref{fig:Gauss_Num_IntGaussRef}.

We show in \Cref{fig:Gauss_Num_IntDim} and
\Cref{fig:Gauss_Num_IntTheta} the dependence of the integrated
autocorrelation times with respectively $p$ in Forward Ref with a
ran-$p$-orthogonal refreshment and with $\theta$ in Forward Ref and
Forward All Ref with a $2$-orthogonal refreshment. The choice of $p$
appears not to be critical and \Cref{fig:Gauss_Num_IntTheta} shows a non-dependence on the angle
$\theta$.

\subsection{ESS for mixture of Gaussian distributions}
\label{sec:ESS_Mix}

We consider the mixture of five Gaussian distributions of
\Cref{sec:mixgaussian}. For small dimensions $d=2,4,5$, the estimated
ESS are similar for all considered schemes, as can be observed in
\Cref{tab:Ran_ESS}.

\newcolumntype{M}{>{\centering\arraybackslash}m{\widthof{RWMF.}}}

\newcolumntype{R}{>{\centering\arraybackslash}m{\widthof{xx}}}
\begin{table}
  \caption{ESS per event for the mixture of five Gaussian distributions. \label{tab:Ran_ESS}}
  \begin{minipage}{\linewidth}
\begin{center}
\vspace{0.5cm}
 \resizebox{\textwidth}{!}{%
 \begin{tabular}{cccccccccccc}
d &h             &\shortstack{\small DF. \\\small No Ref} &\shortstack{\small DF. \\\small Ref} &\shortstack{\small DF. \\\small Ref All}&\shortstack{\small DF. \\\small Full Ref}& \shortstack{\small BPS \\\small No Ref} &\shortstack{\small BPS \\\small Full Ref}&\shortstack{\small IMF \\No Ref}&\shortstack{\small IMF \\\small Ref}&\shortstack{\small RWMF \\\small No Ref} &\shortstack{\small RWMF \\\small Ref}\\\hline

$2$& $x$         &1453\small$\pm$27&--      &--      &1364\small$\pm$2 &1729\small$\pm$34&1608\small$\pm$30&1535\small$\pm$24& --    &1660\small$\pm$25& --\\  

$2$& $\norm{x}^2$&1501\small$\pm$27&--      &--      &1403\small$\pm$22&1726\small$\pm$35&1621\small$\pm$32&1578\small$\pm$25& --     &1696\small$\pm$25& --\\\hline

$4$& $x$         &54\small$\pm$5   &59\small$\pm$4&53\small$\pm$4&55\small$\pm$4   &58\small$\pm$4   &59\small$\pm$4 &62\small$\pm$5   &62\small$\pm$4&58\small$\pm$6   &59\small$\pm$6  \\

$4$& $\norm{x}^2$&63\small$\pm$5   &69\small$\pm$5&60\small$\pm$5&64\small$\pm$4   &67\small$\pm$5   &68\small$\pm$4 &72\small$\pm$6   &72\small$\pm$5&67\small$\pm$6   &67\small$\pm$8\\  \hline

$5$& $x$         &26\small$\pm$3   &25\small$\pm$3&23\small$\pm$5&25\small$\pm$3   &24\small$\pm$4   &26\small$\pm$3  &27\small$\pm$3   &28\small$\pm$3&26\small$\pm$3   &25\small$\pm$3\\

$5$& $\norm{x}^2$&33\small$\pm$4   &32\small$\pm$3&28\small$\pm$4&31\small$\pm$4   &31\small$\pm$5   &33\small$\pm$4&35\small$\pm$4   &35\small$\pm$3&32\small$\pm$3   &31\small$\pm$3 \\\hline
   \end{tabular}
}
\end{center}
\small NOTE:  Results are multiplied by $10^5$. For $d=2$, there is no orthogonal refreshment.
DF.: Direct Forward. IMF: independent-Metropolis Forward. RWMF: random-walk Metropolis Forward.
\end{minipage}
\end{table}

\end{document}